\documentclass[fleqn,usenatbib]{mnras}

\usepackage{newtxtext,newtxmath}
\usepackage[T1]{fontenc}

\DeclareRobustCommand{\VAN}[3]{#2}
\let\VANthebibliography\thebibliography
\def\thebibliography{\DeclareRobustCommand{\VAN}[3]{##3}\VANthebibliography}

\usepackage{graphicx}	
\usepackage{amsmath}	

\usepackage{amssymb}	
\usepackage{url}

\title[Stellar population properties of infalling dEs]{On the accretion of a new group of galaxies onto Virgo: II. The effect of pre-processing on the stellar population content of dEs}

\author[B. Bidaran et al.]{
Bahar Bidaran,$^{1}$\thanks{E-mail: Bidaran@uni-heidelberg.de (BB)}
Francesco La Barbera,$^{2}$
Anna Pasquali,$^{1}$
Reynier Peletier,$^{3}$
Glenn van de Ven,$^{4}$
\newauthor
Eva K. Grebel,$^{1}$
Jesus Falc\'on-Barroso,$^{5,6}$
Agnieszka Sybilska,$^{7}$
Dimitri A. Gadotti$^{8,9}$
Lodovico Coccato,$^{8}$
\\
\\
$^{1}$Astronomisches Rechen-Institut, Zentrum f\"ur Astronomie der Universit\"at Heidelberg, M\"onchhofstra\ss e 12-14, 69120 Heidelberg, Germany\\
$^{2}$INAF – Osservatorio Astronomico di Capodimonte, Salita Moiariello 16, I-80020 Napoli, Italy\\
$^{3}$Kapteyn Astronomical Institute, University of Groningen, Postbus 800, 9700 AV Groningen, the Netherlands\\
$^{4}$Department of Astrophysics, University of Vienna, T\"urkenschanzstrasse 17, 1180 Vienna, Austria\\
$^{5}$Instituto de Astrof\'isica de Canarias, Calle V\'ia L\'actea s/n, E-38205 La Laguna, Tenerife, Spain\\
$^{6}$Departamento de Astrof\'isica, Universidad de La Laguna, E-38200 La Laguna, Tenerife, Spain \\
$^{7}$Sybilla Technologies, Torunska 59, 85-023 Bydgoszcz, Poland\\
$^{8}$European Southern Observatory, Karl-Schwarzschild-Str. 2, D-85748 Garching, Germany\\
$^{9}$Centre for Extragalactic Astronomy, Department of Physics, Durham University, South Road, Durham DH1 3LE, UK\\
}

\date{Accepted XXX. Received YYY; in original form ZZZ}

\pubyear{2021}

\begin{document}
\label{firstpage}
\pagerange{\pageref{firstpage}--\pageref{lastpage}}
\maketitle

\begin{abstract}

Using MUSE spectra, we investigate how pre-processing and accretion onto a galaxy cluster affect the  integrated stellar population properties of dwarf early-type galaxies (dEs). We analyze  a sample of nine dEs  with stellar masses of $\rm \sim 10^9 \, M_\odot$, which were accreted ($\sim$ 2-3 Gyr ago) onto the Virgo cluster as members of a massive galaxy group. We derive their stellar population properties, namely age, metallicity ([M/H]), and the abundance ratio of $\alpha$ elements ([$\alpha$/Fe]), by fitting observed spectral indices with a robust, iterative procedure, and infer their star formation history (SFH) by means of full spectral fitting. We find that these nine dEs are more metal-poor (at the 2-3$\sigma$ level) and significantly more $\alpha$-enhanced than dEs in the Virgo and Coma clusters with similar stellar mass, cluster-centric distance, and infall time. Moreover, for six dEs, we find evidence for a recent episode of star formation during or right after the time of accretion onto Virgo. We interpret the high [$\alpha$/Fe] of our sample of dEs as the result of the previous exposure of these galaxies to an environment hostile to star formation, and/or the putative short burst of star formation they underwent after infall into Virgo. Our results suggest that the stellar population properties of low-mass galaxies may be the result of the combined effect of pre-processing in galaxy groups and environmental processes (such as ram-pressure triggering star formation) acting during the early phases of accretion onto a cluster.
\end{abstract}




\section{Introduction}
Early-type dwarf galaxies (dEs) are low-mass and low-surface brightness systems (with M$_{B}\geq-18$ mag) that are particularly abundant in high-density regions of the Universe, such as galaxy groups and clusters \citep{1974Oemler, 1980Dressler, 1993Whitmore, 1984Postman,2004Kauffmann,2010Peng,2019Davies}. These galaxies are generally dominated by metal-poor stellar populations (with [M/H]$<$0.00 [dex]) that cover a rather wide range of light-weighted ages (3$<$Age [Gyr]$<$14). Most of the dEs observed to date are dominated by solar-scaled [$\alpha$/Fe] abundance ratios \citep{2003Geha,2004vanZee,2010Paudel,2018Sen}. The level of [$\alpha$/Fe] enhancement in dEs is similar to the high-metallicity stars of the Milky Way (MW) and the Large Magellanic Cloud (LMC) \citep{2018Sybilska}.

Dwarf elliptical galaxies show diverse morphologies, dynamics, and stellar population properties. For instance, in the core of clusters, 
dEs are often old, metal-poor, and slow-rotating systems 
\citep[e.g.,][]{2006Geha, 2009Koleva, 2011Toloba, 2014Toloba, 2014Rys, 2017Sybilska,2020Scott}. In the Lambda cold dark matter ($\Lambda$CDM) framework \citep{2000Cole} of hierarchical structure formation, dEs might have formed
from proto$-$galaxies in the early Universe and ever since evolved passively \citep[e.g.,][]{2003DeRijcke, 2017Wheeler}.
Nonetheless, the presence of disc features \citep{2006Lisker,2006LiskerII}, stripped tails of HI gas, and prolonged star formation history akin to that of low-mass late-type star-forming galaxies, challenge the picture presented above. Therefore, \cite{1985Kormendy} suggested the transformation of late-type star-forming galaxies under environmental effects as another possible formation channel for part of the present-day dEs in clusters and galaxy groups \citep[also check  ][]{1988Binggeli,2013Lisker,2014Boselli}. 

Within a massive host halo, ram pressure stripping (RPS) exerted by the intracluster medium can effectively exhaust the star formation activity of satellites by depleting their cold gas reservoir over a short time-scale of $\sim$ 1 Gyr \citep{1972Gunn,2010Hester,2011Fumagalli,2014Kenney,2021Boselli,2021Roberts}. 
Ram pressure
can also trigger local star formation enhancements, as a result of gas compression in the leading part and core of infalling satellites \citep[e.g.,][]{2017Lee,2018Fossati,2021Boselli}\citep[but see:][]{2021Mun}. 
Furthermore, the gravitational potential well of the host halo and multiple close encounters with other halo members apply tidal forces on satellites over a time-scale of several Gyr, and are thus able to modify their dynamics, morphology, and star formation history. These tidal interactions are known as galaxy strangulation \citep{1980Larson} and harassment, respectively \citep{1996Moore,1998Moore,2006Boselli}. Hence we expect to find correlations between the properties of dEs (i.e., their morphologies, dynamics, colors and stellar populations) and their local density, host halo mass, and cluster-centric distance. 

Observations have shown that red and quenched dEs make up a large galaxy population in clusters and massive galaxy groups \citep{2008Chilingarian,2009Smith,2013Lisker,2017Sybilska,2021Janz, 2021Venhola,2021Su}. The fraction of old and quenched low-mass galaxies increases with the host halo mass, as observed by \cite{2010Pasquali} and \cite{2021Gallazzi}. Within a massive host halo, however, their relative number decreases toward larger clustercentric distances \citep{2008Chilingarian,2009Smith,2013Lisker}. For instance, in the Coma cluster, \cite{2009Smith} showed that the population of young ($\sim$ 3 Gyr) dEs with solar-scaled [Mg/Fe] abundance ratios increases toward the cluster outskirts. They also found that in the core of Coma, dEs are mostly old ($\sim$ 10 Gyr) and Mg-enhanced. A similar behavior is also observed in Abell 496 by \cite{2008Chilingarian} and in Virgo by \cite{2016Liu}. Such observed trends emphasize the role of environment in transforming its low-mass satellite galaxies.

\setlength{\tabcolsep}{3.pt}
\begin{table*}
\caption{\label{BL_sample} Properties of our dEs}
\centering
\begin{tabular}{c c l l c c c c c c c c}
\hline
Object      & type    & $\alpha$ (J2000)    &   $\delta$ (J2000)  &  \textit{z}$^{a}$  & \textit{R}$_{\rm e}^{a}$ \rm [arcsec] & \textit{M}$_{\rm r}^{a}$ \rm [mag] & \textit{g-r}$^{a}$ \rm [mag] &  $ A_{\rm V}^{b}$ [mag]   & $\sigma_{\rm Re}$ $^{\rm c}$ [kms$^{-1}$] & $\lambda_{\rm Re}$ $^{\rm c}$& TET [hour]\\
\hline
\hline
VCC 0170 & dE(bc) & 12 15 56.30 & +14\ 25\ 59.2 & 0.00472 & $31\farcs57$ & $-$17.62 & 0.59  & 0.089& 27.1 $\pm$ 10.6 & 0.45 $\pm$ 0.03 & 4\\
VCC 0407 & dE(di) & 12 20 18.80 & +09\ 32\ 43.1 & 0.00626 & $18\farcs38$ & $-$17.37 & 0.61  & 0.057 & 32.3 $\pm$ 8.8 & 0.67 $\pm$ 0.03 & 2\\
VCC 0608 & dE(di) & 12 23 01.70 & +15\ 54\ 20.2 & 0.00607 & $25\farcs77$ & $-$17.58 & 0.60  & 0.072 & 25.1 $\pm$ 9.2 & 0.38 $\pm$ 0.04 & 5\\
VCC 0794 & dE(nN) & 12 25 21.61 & +16\ 25\ 46.9 & 0.00558 & $37\farcs33$ & $-$17.29 & 0.61  & 0.065 & 33.0 $\pm$ 7.5 & 0.48 $\pm$ 0.04 & 3\\
VCC 0990 & dE(di) & 12 27 16.93 & +16\ 01\ 28.1 & 0.00573 & $10\farcs31$ & $-$17.43 & 0.62  & 0.080 & 36.0 $\pm$ 5.6 & 0.27 $\pm$ 0.03 & 3\\
VCC 1833 & ---    & 12 40 19.70 & +15\ 56\ 07.1 & 0.00569 & $8\farcs52$  & $-$17.44 & 0.61  & 0.099 & 34.4 $\pm$ 6.0 & 0.15 $\pm$ 0.06 & 1.5\\
VCC 1836 & dE(di) & 12 40 19.50 & +14\ 42\ 54.0 & 0.00668 & $42\farcs27$ & $-$17.45 & 0.58  & 0.079 & 38.5 $\pm$ 8.2 & 0.55 $\pm$ 0.03 & 4\\ 
VCC 1896 & dE(di) & 12 41 54.60 & +09\ 35\ 04.9 & 0.00629 & $14\farcs98$ & $-$17.04 & 0.62  & 0.047 & 27.0 $\pm$ 7.2 & 0.22 $\pm$ 0.04 & 3\\
VCC 2019 & dE(di) & 12 45 20.40 & +13\ 41\ 34.1 & 0.00607 & $18\farcs60$ & $-$17.65 & 0.63  & 0.060 & 31.2 $\pm$ 6.5 & 0.73 $\pm$ 0.05 & 2\\
\hline
\end{tabular}\\
The columns show: Name of target, morphological type, right ascension and declination, redshift (\textit{z}), effective radius ($R_e$) measured at the half-light major axis, absolute \textit{r}-band magnitude ($M_r$), \textit{g-r} colour measured at 1$R_{\rm e}$, foreground Galactic extinction in the V-band (A$_{V}$), stellar velocity dispersion at 1 $R_{e}$ ($\sigma_{\rm Re}$), specific angular momentum at 1$R_{e}$ ($\lambda_{\rm Re}$), total exposure time (TET). \\
a:\,\cite{2006Lisker},
b:\,\cite{2011Schlafly},
c:\,\cite{2020Bidaran}
\\
\end{table*}

Nonetheless, such a transformation may start long before dEs are accreted onto their current host halo. According to the $\Lambda$CDM framework, clusters grow through the accretion of galaxies, either individually or in pairs and more populated groups where they may have already experienced environmental effects. This is known as pre-processing \citep{2004Mihos,2004Fujita} and can for example  explain the presence of gas-deprived galaxies or galaxies with a low star-formation rate, as well quenched galaxies beyond the cluster virial radius \citep{2001Kodama,2002Lewis,2012Mahajan,2015Haines,2021Donnari} 
Distinguishing between the effects of pre-processing and the effects of the current host halo is, however, not straightforward, since their differences dilute through time. One possible way to disentangle current halo processing from pre-processing can
be the study of newly accreted dEs in dynamically young galaxy clusters like Virgo. 

In the projected phase-space diagram of $\sim$ 620 Virgo galaxies, \cite{Lisker2018} discovered a sample of nine dEs (with $-$17$\geq$M$_{r}$ [mag]>$-$18) that according to N-body simulations of \cite{2015Vijayaraghavan} should have recently been accreted onto Virgo as gravitationally bound members of a massive galaxy group (with M$_{\star}$/ M$_{\odot}$ $\sim$ 10$^{13}$). According to the simulations, this infall occurred along the observer's line of sight about 2-3 Gyr ago. In \cite{2020Bidaran} (hereafter B20), we confirmed this relatively recent infall time using the results of \cite{2019Pasquali} and \cite{2019Smith} who dissected the projected phase-space diagram into several zones of different average infall time. In B20, we conducted a kinematic analysis of these nine dEs based on MUSE (Multi-Unit Spectroscopic Explorer) data and showed that, despite their similar stellar mass range and infall time to Virgo, they feature diverse kinematic properties. In particular, we showed that their specific angular momentum ($\lambda_{R}$) profiles are intermediate between equally massive star-forming galaxies in the field and equally-massive Virgo dEs with earlier infall times. We interpreted this diversity in their $\lambda_{R}$ profiles as possible footprints of pre-processing in their previous group environment. On these grounds, we would expect to detect the effects of such pre-processing also in their stellar population properties.

In this study, we use the MUSE spectra of B20 to derive the integrated stellar population properties and star formation history of our sample of dEs. In order to disentangle group pre-processing from cluster early processing, we also compare our sample with other Virgo and Coma dEs within a similar stellar mass range and infall time. The latter is estimated using the position of cluster dEs in the projected phase-space diagram of their host halos \citep{2019Pasquali, 2019Smith}. The results of this investigation can also help to understand how dense environments, such as the Virgo cluster, alter the stellar population properties of their satellites during their accretion event and early phases of infall. 

This paper is organized as follows: In Section \ref{Data} we briefly introduce our main sample and the comparison samples that we use in our analysis. In Section \ref{Analysis} we describe the methods that we utilize for deriving stellar population parameters from the MUSE cubes of our main sample. In Section \ref{Results} we present the integrated stellar population properties and star formation histories of our sample, and compare them with those of the comparison samples. We discuss our results in Section \ref{Discussion} and summarize them in Section \ref{Conclusion}.

\section{Data}\label{Data}
\subsection{Our sample}
We analyze the integrated stellar population properties for the sample of nine Virgo dEs (-17$\geq$M$_{r}$ [mag]>-18) studied by B20, where their selection criteria are presented and the data reduction is described. Briefly, we observed these galaxies using the MUSE instrument mounted on the Very Large Telescope (VLT), following a science verification proposal in the period of December 2016 to February 2017, and February 2018 to July 2018 (P98, ESO programs 098.B-0619 and 0100.B-0573; PI: Lisker). Our MUSE dataset maps a field of view of 1$\times$1 arcmin$^{2}$ with a spatial resolution of 0.2 arcsec/pixel. The spectral resolution of MUSE is wavelength dependent and its average full width at half maximum (FWHM) is 2.5 \AA\ \citep{2010Bacon}. MUSE spectra cover the optical wavelength range from 4750 to 9350 \AA. All dEs in our sample were observed with a nearly constant seeing (mean FWHM $\sim$ 1.6'').

We summarize the properties of our sample of dEs in Table \ref{BL_sample} where we list the name of our target, their morphological type, coordinates, redshift, $r$-band effective radius, absolute $r$-band magnitude, $g-r$ colour, foreground Galactic extinction in the V-band (A$_{V}$), stellar velocity dispersion at 1 effective radius (R$_{e}$), specific angular momentum at 1R$_{e}$ ($\lambda_{\rm Re}$), and total exposure time (TET) of the MUSE observations. All the reported values, except for A$_{V}$, $\lambda_{\rm Re}$ and $\sigma_{\rm Re}$, are taken from \cite{2006Lisker}. For each dE, the A$_{V}$ value is taken from \cite{2011Schlafly} through the NASA/IPAC Extragalactic Database (NED). The values of $\lambda_{\rm Re}$ and $\sigma_{\rm Re}$ for each dE are taken from B20. The absolute r-band magnitudes and colours were corrected for Galactic foreground extinction by \cite{2006Lisker} and \cite{2008Janz,2009Janz}. As estimated in B20, our dEs fall in the stellar mass range of 8.9 < log (M$_\star$ [M$_\odot$]) < 9.2. 

\subsection{The comparison samples}
In this study, we compare our results with those for other dEs in the Virgo and Coma clusters with -17 $\geq$ M$_{r}$ [mag]$\ge$ -18 \footnote{We assume a distance of 16.5 Mpc to the Virgo cluster \citep[e.g.,][]{2007Mei} and 100 Mpc to the Coma cluster \citep[e.g.,][]{2008Carter}.}. This cut in absolute magnitude corresponds to a narrow mass range that allows us to investigate environmental effects at fixed stellar mass.
The comparison samples include the following dEs from the literature:
\begin{itemize}
\item 8 Virgo dEs from \cite{2014Toloba} and 6 Virgo dEs from \cite{2018Sen}. The stellar population properties of these galaxies were obtained from long-slit spectra (observed with three different telescopes: William Herschel Telescope, Isaac Newton Telescope, and VLT using the FORS2 spectrograph). 
\item 4 Virgo dEs from \cite{2010Paudel}. Their stellar population properties were derived using long-slit spectra acquired with the FORS2 spectrograph on the VLT. 
\item 13 Virgo dEs from \cite{2017Sybilska}. The stellar population properties of these galaxies were measured using IFU spectra obtained with the SAURON spectrograph at the William Herschel Telescope. 
\item 47 Coma dEs from \cite{2009Smith}. Their stellar population properties were determined using long-slit spectra from the Hectospec fiber-fed spectrograph at the MMT telescope.

The addition of the Coma dEs to the Virgo comparison sample compensates for the fact that measurements of the [$\alpha$/Fe] ratio for Virgo dEs are quite scarce. Moreover, it also can remedy possible biases due to Virgo being a dynamically young galaxy cluster with substructures, which may not fully show up in the projected phase-space diagram as we define it. Thus, adding the Coma cluster, which is more relaxed than Virgo, can be beneficial for generalizing observed trends in our study.

\end{itemize}

Throughout this paper, we refer to the dEs of the comparison samples that belong to the Coma and Virgo clusters as the “Coma dEs” and “Virgo dEs”, respectively. All dEs in both clusters and including our sample dEs are referred to as the “cluster dEs”, whenever needed. 

\section{Analysis}\label{Analysis}
\subsection{Setup and pre-processing of the data}\label{Analysis: setup}
 
We mask the non-related background galaxies and foreground stars in each dE MUSE datacube. We then construct the integrated MUSE spectrum by averaging each dE datacube within 1R$_{e}$, where spaxels with SNR$<$3 have been discarded. We use the Galactic extinction law of \cite{1989Cardelli} to correct these integrated spectra for Galactic foreground extinction. Here we adopt R$_{v}$ = 3.10 and the A$_{V}$ values as reported in Table \ref{BL_sample}. Additionally, we correct the spectra for possible nebular emission lines. As discussed in B20, we have detected relatively strong nebular emission lines (i.e., H$\beta$, H$\alpha$, [OIII], [NII], and [SII]) in the central regions of VCC0170. We perform the emission line correction for all dEs in our MUSE sample, using the GANDALF \citep[Gas AND Absorption Line Fitting;][]{2006Sarzi, 2006MNRAS.369..529F} package. This software treats each emission line as a Gaussian function, and simultaneously fits the stellar continuum and emission lines. The observed spectrum is corrected by subtracting the resulting residual emission line spectrum obtained for the best fit. We use these corrected integrated spectra to measure line indices, and later to determine the stellar population properties, as well as star formation histories of our dEs.

In this study, we use single stellar population (SSP) models of \cite{2010Vazdekis,2015Vazdekis} based on the MILES stellar library \citep{2006MNRAS.371..703S,2007Cenarro,2011A&A...532A..95F}. These SSP models span the age range of 0.5 to 14.0 Gyr, the metallicity range of -1.26 to 0.06 dex, and the $[\alpha$/Fe] values of 0.00 dex (solar-scaled models) and 0.40 dex ($\alpha$-enhanced models). The SSPs were constructed using BASTI isochrones \citep{2004Pietrinferni} and a bi-modal initial mass function (IMF) with a slope of 1.3 \citep{1996Vazdekis}. We broaden them from their original spectral resolution (FWHM = 2.51 \AA) to FWHM = 5 \AA, i.e. the LIS-5 \AA\ system \citep{2010Vazdekis}, which is consistent with the typical velocity dispersion of dEs.

\begin{figure}
\includegraphics[scale=0.6]{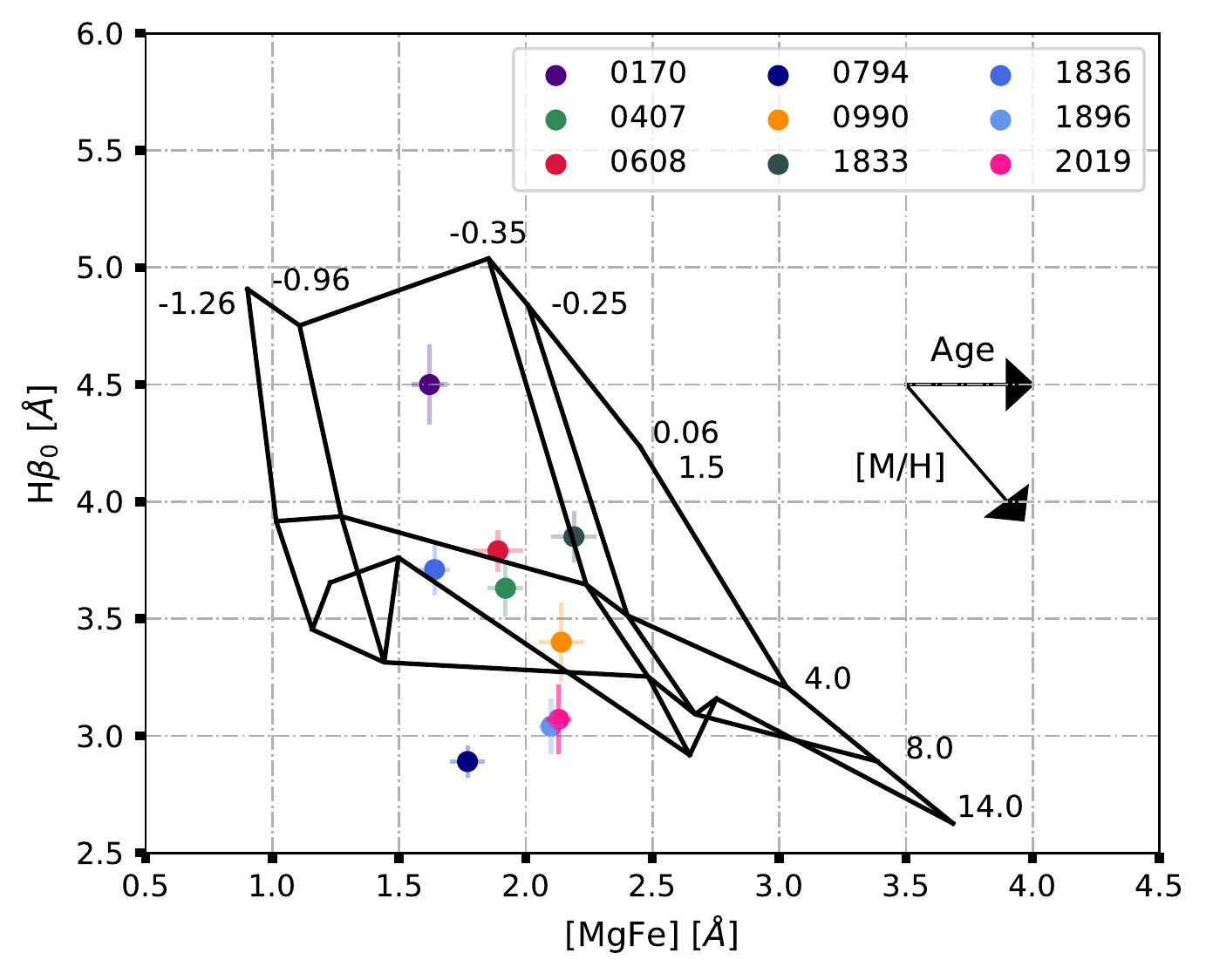}
\includegraphics[scale=0.6]{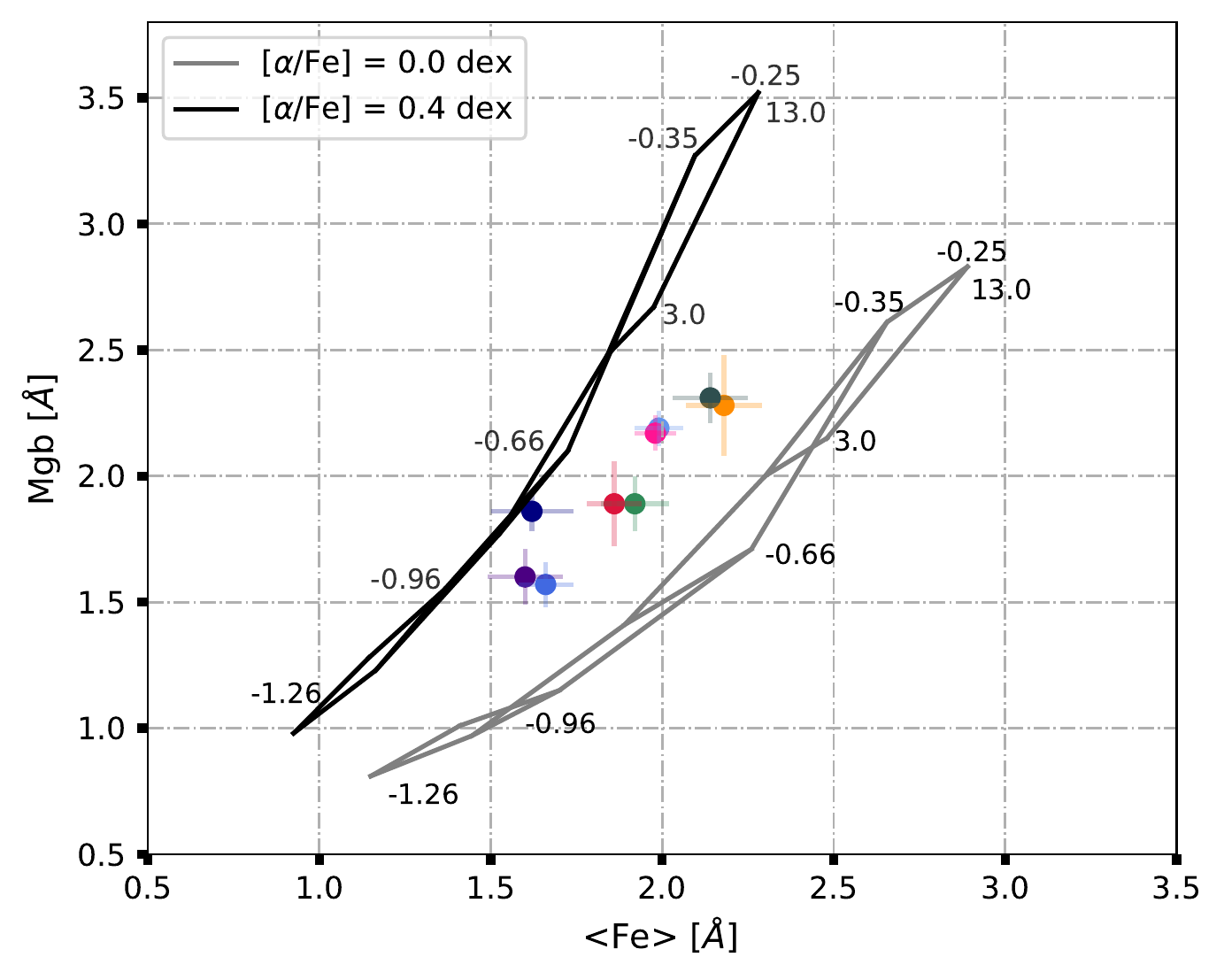}
\caption{\textit{Top panel:} The H$\beta_{0}$ vs. [MgFe] grid. SSP model predictions with [$\alpha$/Fe] = 0.4 dex are plotted with black solid lines. Following the horizontal arrow from left to right, metallicity increases along the grid from -1.26 to +0.06 dex. Similarly, as shown by the diagonal arrow and from top to bottom, age increases from 1.5 to 14 Gyr}. \textit{Lower panel: The Mgb vs. $<\rm Fe>$ grid. Model predictions with [$\alpha$/Fe]=0.00 and 0.4 dex for two ages (3.0 and 13 Gyr) and the metallicity range of [-1.26, -0.96, -0.66, -0.35, -0.25] dex are shown with gray and black solid lines, respectively. In both panels, our dEs are denoted with colour-coded circles as listed in the legend in the top panel.}
\label{Grids}
\end{figure}

\setlength{\tabcolsep}{3.pt}
\begin{table*}
\caption{\label{Integrated values} Measured indices and SSP-equivalent properties of our dEs within 1$R_{e}$}
\centering
\resizebox{\textwidth}{!}{\begin{tabular}{c c c c c c c c c c c c}
\hline
Object      & H$\beta$    & H$\beta_{0}$    &  Mgb    & Fe5015   & Fe5270 & Fe5335 & [MgFe] & $<\rm Fe>$ & Age &  [M/H] & [$\alpha$/Fe]\\
            &  (\AA)      &(\AA)            &(\AA)    &(\AA)     &(\AA)   &(\AA)   &(\AA) &(\AA) & (Gyr) & (dex) & (dex)\\
\hline
\hline
VCC 0170 & 3.39 $\pm$ 0.08   & 4.50 $\pm$ 0.17    & 1.60 $\pm$ 0.11  &    2.75 $\pm$ 0.25 & 1.71 $\pm$ 0.09 & 1.49 $\pm$ 0.21 & 1.62 $\pm$ 0.07 & 1.60 $\pm$ 0.11 &  2.10$^{+0.16}_{-0.38}$  &    $-$0.61$^{+0.04}_{-0.04}$ & 0.26$^{+0.09}_{-0.08}$ \\[0.2cm]
VCC 0407 & 2.70 $\pm$ 0.07 & 3.63 $\pm$ 0.12 & 1.89 $\pm$ 0.11  & 3.59 $\pm$ 0.22   & 2.01 $\pm$ 0.11   & 1.83 $\pm$ 0.18 & 1.92 $\pm$ 0.07 & 1.92 $\pm$ 0.10& 5.78$^{+7.34}_{-1.02}$ & $-$0.59$^{+0.06}_{-0.06}$ & 0.20$^{+0.08}_{-0.08}$\\[0.2cm]
VCC 0608 & 2.79 $\pm$ 0.14 & 3.79 $\pm$ 0.09  & 1.89 $\pm$ 0.17  & 3.58 $\pm$ 0.23  & 1.95 $\pm$ 0.13  & 1.77 $\pm$ 0.09   & 1.89 $\pm$ 0.10 & 1.86 $\pm$ 0.08& 4.55$^{+1.06}_{-0.53}$& $-$0.57$^{+0.06}_{-0.08}$& 0.22$^{+0.08}_{-0.08}$\\[0.2cm]
VCC 0794 & 2.12 $\pm$ 0.09  &   2.89 $\pm$ 0.07   &  1.86 $\pm$ 0.08 & 2.81 $\pm$ 0.18 & 1.75 $\pm$ 0.15   &   1.50 $\pm$ 0.20  & 1.77 $\pm$ 0.07 & 1.62 $\pm$ 0.12& 9.98$^{+0.31}_{-0.40}$ & $-$0.73$^{+0.02}_{-0.02}$& 0.43$^{+0.06}_{-0.08}$\\[0.2cm]
VCC 0990 & 2.37 $\pm$ 0.12 &  3.40 $\pm$ 0.17 & 2.28 $\pm$ 0.20 & 4.17 $\pm$ 0.20 & 2.29 $\pm$ 0.12  &  2.00 $\pm$ 0.09 & 2.14 $\pm$ 0.09 & 2.18 $\pm$ 0.11 & 7.77$^{+3.59}_{-1.59}$& $-$0.53$^{+0.06}_{-0.06}$& 0.14$^{+0.08}_{-0.08}$\\[0.2cm]
VCC 1833 & 2.78 $\pm$ 0.10 & 3.85 $\pm$ 0.11 & 2.31 $\pm$ 0.10   & 4.36 $\pm$ 0.24   & 2.30 $\pm$ 0.13   & 2.00 $\pm$ 0.10  & 2.19 $\pm$ 0.09 & 2.14 $\pm$ 0.11& 3.66$^{+1.28}_{-1.06}$& $-$0.37$^{+0.12}_{-0.08}$& 0.19$^{+0.08}_{-0.08}$\\[0.2cm]
VCC 1836 & 2.81 $\pm$ 0.14 & 3.71 $\pm$ 0.11    & 1.57 $\pm$ 0.09   & 3.14 $\pm$ 0.14 & 1.77 $\pm$ 0.09   & 1.56 $\pm$ 0.13    & 1.64 $\pm$ 0.06 & 1.66 $\pm$ 0.08& 6.84$^{+5.96}_{-2.25}$& $-$0.77$^{+0.08}_{-0.06}$& 0.28$^{+0.04}_{-0.06}$\\[0.2cm]
VCC 1896 & 2.16 $\pm$ 0.09  & 3.04 $\pm$ 0.12   & 2.19 $\pm$ 0.07  & 3.55 $\pm$ 0.16 & 2.04 $\pm$ 0.10  & 1.95 $\pm$ 0.09   & 2.10 $\pm$ 0.05 & 1.99 $\pm$ 0.07 & 10.50$^{+0.48}_{-0.53}$ & $-$0.59$^{+0.02}_{-0.02}$& 0.26$^{+0.04}_{-0.06}$\\[0.2cm]
VCC 2019 & 2.34 $\pm$ 0.09 & 3.07 $\pm$ 0.15    & 2.17 $\pm$ 0.07   & 3.71 $\pm$ 0.14  & 2.09 $\pm$ 0.10  & 1.87 $\pm$ 0.07  & 2.10 $\pm$ 0.05 & 1.98 $\pm$ 0.06 & 10.46$^{+0.48}_{-1.02}$ & $-$0.59$^{+0.04}_{-0.02}$& 0.26$^{+0.04}_{-0.04}$\\[0.1cm]
\hline
\end{tabular}}\\
The columns show: (1) Target name, (2 -9) measured indices and their corresponding errors, (10-12) light-weighted age, metallicity, and [$\alpha$/Fe]. All the values are measured using  the dEs' integrated MUSE spectra. 
\\

\end{table*}

\subsection{Line-strength measurements} \label{Line-Strength indices}
To derive SSP-equivalent stellar ages, metallicities, and [$\alpha$/Fe] ratios, we use
the set of line indices 
defined in Table \ref{Index banpass}, where the H$\beta$ and H$\beta_0$ indices are age-sensitive, and the Fe5015, Mgb5177, Fe5270, and Fe5335 indices are sensitive to metallicity. The indices are measured after smoothing all observed spectra to the 5 \AA\ resolution of the LIS-5\AA\ system \citep{2010Vazdekis}. 
The convolution is based on a Gaussian kernel whose FWHM$_{\rm final}$ is defined as: 

\begin{equation}
    \rm FWHM_{\rm final} = \sqrt{\rm FWHM_{\rm LIS5.00}^{2} - (\rm FWHM_{\rm MUSE}^{2} + \rm FWHM_{\rm \sigma}^{2})}
\end{equation}

\noindent where FWHM$_{\rm MUSE}$ is the wavelength-dependent MUSE resolution \citep{2017Bacon} and FWHM$_{\rm \sigma}$ corresponds to the stellar velocity dispersion of each dE as reported in Table \ref{BL_sample}. 

To estimate the random error on each measured index, we consider three possible uncertainty sources: the error on the stellar velocity dispersion (affecting the absorption line width), the error on the redshift estimate (affecting the accurate placement of the index band-passes), and the error on the spectral flux \citep{2006Kuntschner}. Thus, we perturb the flux values in the spectra using a Gaussian distribution with width equal to the estimated flux errors. Furthermore, we randomly shift the perturbed spectra using a normal Gaussian distribution of redshift errors. Additionally, in each perturbation, we change the convolution kernel size by randomly shifting the stellar velocity dispersion within a normal distribution constructed on the corresponding $\sigma$ errors. Based on this error treatment, we run 125 Monte Carlo (MC) iterations for each spectrum and assume the standard deviation of the multiple measurements of each index as its corresponding error. Furthermore, we notice that since stellar population models have their own uncertainties, matching observed and model line-strengths for high S/N ratio spectra may lead to unrealistically small errors on stellar population parameters. Indeed, using spectral fitting (with the software STARLIGHT), we estimated that the quality of the fits to our MUSE spectra does not improve significantly when S/N is larger than $\sim 150$, corresponding to a typical uncertainty of $\sim 0.05$~\AA\ for the line-strengths we analyze in the present work. Therefore, we decided to add a minimum uncertainty of 0.05~\AA\ in quadrature to the errors on observed line-strengths.  The measured indices and their uncertainties for our sample of dEs are reported in Table \ref{Integrated values}.

In the top panel of Fig.\ref{Grids}, we compare the H$\beta_{0}$ and [MgFe] indices measured from the integrated spectrum of each dE (colour-coded circles) with SSP predictions (black grid, corresponding to SSPs with [$\alpha$/Fe] = 0.40 dex). 
The metallicity-sensitive and composite [MgFe] index is derived as in  \cite{2003Thomas}: 
\begin{equation}
[\rm MgFe] =  \sqrt{Mgb \times (0.72 \times Fe5270 + 0.28 \times Fe5335)}
\end{equation}
The diagonal lines of the grid have constant metallicity (from left to right: -1.26, -0.96, -0.35, -0.25, and 0.06 dex), while the horizontal lines have constant age (from top to bottom: 1.5, 4.0, 8.0, and 14 Gyr). The distribution of our dEs in H$\beta_{0}$ vs [MgFe] indicate that they span quite a range of ages.

We plot the Mgb and $<$Fe$>$ indices of our dEs (with $<$Fe$>$ being the average of Fe5270 and Fe5335) in the bottom panel of Fig.\ref{Grids}, together with two
SSP grids corresponding to [$\alpha$/Fe] = 0.00 dex and 0.40 dex, in gray and black, respectively. Each grid shows the SSP predictions for the ages of 3.0 and 13.0 Gyr and the metallicity range of [-1.26, -0.96, -0.66, -0.35, and -0.25] dex. Note that in this work, we use Mgb5177 as a proxy for $\alpha$ abundance. We notice that our dEs follow a quite tight distribution on this plane, being systematically and consistently $\alpha$-enriched.

\subsection{SSP-equivalent properties}\label{Method: fitting}

In order to fit the observed indices with those provided by the SSP models,
we construct finer grids of SSP predictions, by linearly interpolating the SSP models in age-[M/H]-[$\alpha$/Fe]-[index] space, using steps of 0.02 dex and 0.015 dex in metallicity and [$\alpha$/Fe], respectively. For what concerns the interpolation in age, one should consider that variations in age are much larger for young ages (0.5 $<$ age [Gyr]$<$ 3) relative to old ones (3 $<$ age [Gyr] $<$ 14). Hence, to keep the number density of models approximately constant over the SSPs' age range, we adopt age steps of 0.01 Gyr and 0.04 Gyr for young and old ages, respectively. As shown in the bottom panel of Fig.\ref{Grids}, a couple of galaxies are quite close to the black grid of SSPs with [$\alpha$/Fe] = 0.4 dex, the maximum value for which models were computed.
This may raise concerns regarding false parameter estimates and systematic errors. Moreover, the [$\alpha$/Fe] of Coma and Virgo dEs from \cite{2009Smith} and \cite{2017Sybilska} span a range of $\sim$[-0.3,0.6] dex. Therefore, we decide to also extrapolate the SSP models in steps of 0.015 dex to fill a final [$\alpha$/Fe] range of -0.3 dex to 0.7 dex.

To translate the measured indices into model-predicted stellar population properties, we employ a $\chi^{2}$ minimization approach in four steps, each time fitting over certain index pairs:

\textbf{Step 1:} we construct the finer H$\beta_{0}$-[MgFe] grid for solar-scaled SSP models. Based on the galaxy position in this grid, we fit the SSP-equivalent age by running the $\chi^{2}$ minimization method. We construct the probability distribution function (PDF) of age in this step by repeating the fitting procedure 25 times in an MC framework. In each MC run, we shift the measured indices accounting for their errors. The resulting age PDF is used in the next step.

\textbf{Step 2:} we construct the finer Mgb-<Fe> grid using models at values randomly selected from the age PDF delivered by Step 1. Then, we measure the [$\alpha$/Fe] PDF from 100 MC realizations, each based on the $\chi^{2}$ minimization method. We repeat this test for 50 randomly selected values from the age PDF. The final [$\alpha$/Fe] PDF, which is passed to Step 3, is the average of all these 50 random selections.

\textbf{Step 3:} we randomly select 100 values from the [$\alpha$/Fe] PDF delivered by Step 2, and execute 200 MC iterations for each selected value. Here, we revisit the finer H$\beta_{0}$-[MgFe] grid, each time constructed for a randomly selected [$\alpha$/Fe] value from the [$\alpha$/Fe] PDF of Step 2. The reason why we repeat this step (with respect to Step 1), is to account for the dependence of H$\beta_{0}$ on [$\alpha$/Fe] \citep[see][]{2015Vazdekis}. The final age (in Gyr) and [M/H] (in dex) are measured as the median of their corresponding PDFs constructed by the end of Step 3.

\textbf{Step 4:} as the last step, we repeat the same methodology of Step 2 by randomly selecting 100 ages from the age PDF of Step 3 and executing 200 MC iterations for each selected age. The final [$\alpha$/Fe] value (in dex) is measured as the median of the resulting PDF in this step. 

In our iterative fitting approach, each stellar population parameter is derived based on the absorption features to which it is more sensitive to. An alternative approach would be performing simultaneous fits over the parameter space, where a combination of absorption features with different levels of sensitivity is taken into account. Here, we choose to apply an iterative index fitting approach to gain more robust results. In the last three columns of Table \ref{Integrated values}, we report the integrated stellar population parameters of our dEs, measured as medians of PDFs from Steps 3 and 4. The reported errors are the 84th and 16th percentiles of the final PDFs. It should be noted that the stellar population properties in Table \ref{Integrated values} are light-weighted values. Since the recent formation of even a few young and hot stars in galaxies can affect their composite light drastically, our measurements might be biased toward the recent star formation in these galaxies, if any. However, the effect is not expected to be severe for [M/H] results, as the contribution of hot and young stars to metal lines of the integrated spectra is not significant \citep{2005Trager, 2007Serra}.

\subsection{Full spectrum fitting using STARLIGHT}\label{method:STARLIGHT}
To investigate the star formation histories of our sample of dEs, we utilize the publicly available full-spectrum fitting routine STARLIGHT\footnote{\url{http://www.starlight.ufsc.br/}}. STARLIGHT constructs the best-fitting synthetic spectrum by linearly combining a set of SSP models with different ages and metallicities \citep{2005CidFernandes}. We perform the full spectrum fitting over the integrated spectra of our dEs to measure their light- and mass-weighted stellar population properties. Since full-spectrum fitting does not consider a single SSP (as in our index fitting approach), but a combination of SSP models, it provides an additional and complementary estimate of the stellar populations content of our dEs. Furthermore, we use the STARLIGHT population vectors (the fraction of either light or mass contributed by each SSP to the best-fitting synthetic spectrum) to build the star formation histories of our dEs. 

We perform the fitting over the spectral range of 4750 \AA\  to 5540 \AA. This wavelength range contains the most crucial absorption features that are sensitive to age and [M/H] and is less affected by telluric lines \citep[for more explanation on possible impacts of selected wavelength ranges on the results of full-spectrum fitting, see:][]{2020STARLIGHTperform}. To normalize the input flux, STARLIGHT uses the median of the observed spectrum in a specific user-defined wavelength range, called the "S/N window". We define our preferred S/N window in the wavelength range of 4750 \AA\ to 4800 \AA. 

In order to compare the results of STARLIGHT with those of the Lick indices fitting, we run STARLIGHT using the same set of SSP models as for index fitting (i.e. same IMF and isochrones, see Section \ref{Analysis: setup}). In STARLIGHT, we use SSP models in the age range of 1 to 14 Gyr (with $\Delta_{\rm age}$ = 1 Gyr) and the metallicity range of -1.26 to -0.25 dex. As explained in Section \ref{Method: fitting}, absorption features that are present in our adopted fitting wavelength range are not only age and [M/H] dependent but also are sensitive to [$\alpha$/Fe]. As discussed in \cite{2015Vazdekis}, the full spectrum fitting does not correctly estimate [$\alpha$/Fe] ratios, unless a narrow wavelength range around the Mgb absorption feature is selected. On the other hand, STARLIGHT only accepts a limited number of SSP models as its defined stellar base (up to 300). Thus, we adopt a simplified approach where [$\alpha$/Fe] is fixed to the value derived with the index fitting method (i.e., values in Table \ref{Integrated values}).

\begin{figure*}
    \centering
	\includegraphics[scale=0.60]{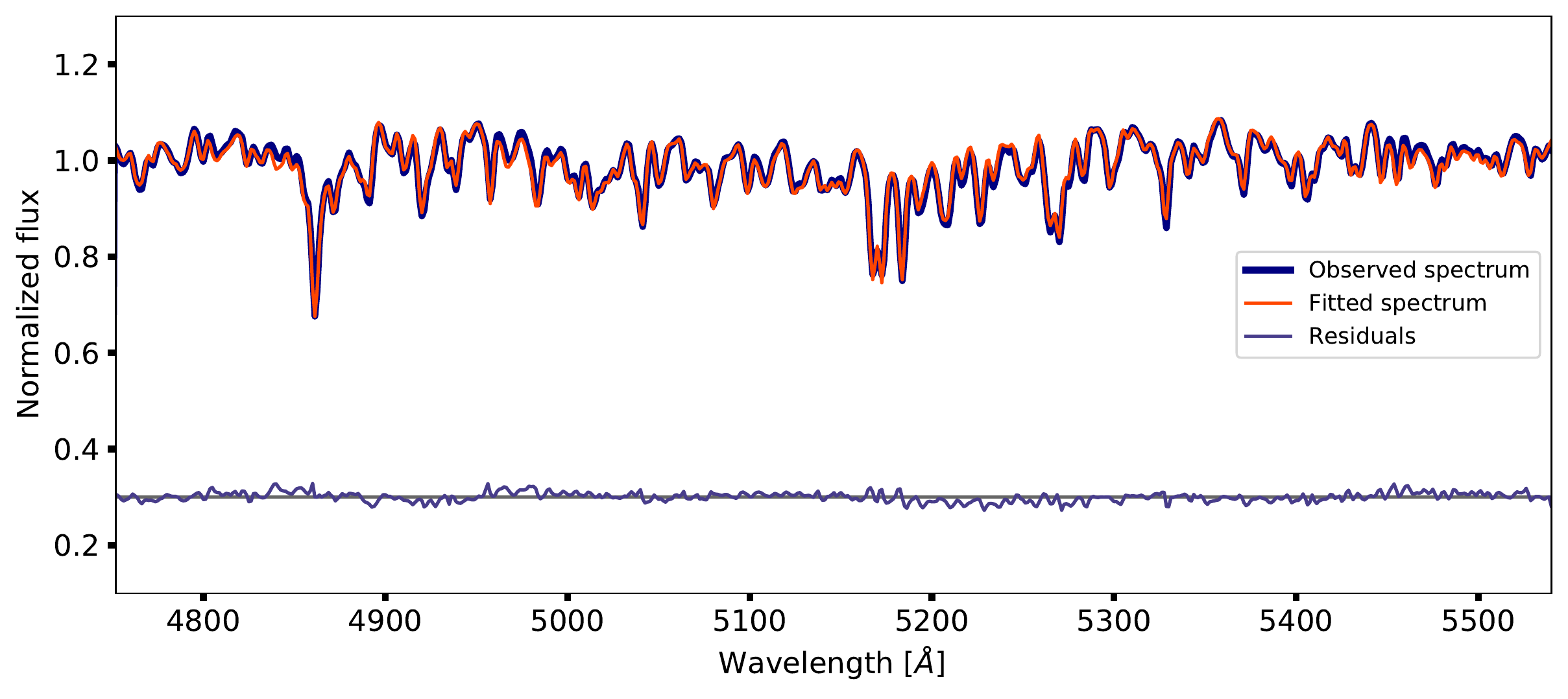}
	\caption{An example of the full spectrum fitting over the averaged spectrum of VCC2019. The observed and normalized spectrum is plotted in dark blue. STARLIGHT best-fitting synthetic spectrum is shown in red, and the residuals of the fit are shown in light purple. For better legibility, the residuals are shifted up by 0.3. } 
	\label{STARLIGHTFIT_2019}
\end{figure*}

To do so, we linearly interpolate the flux of the SSP models with similar age and [M/H], but different [$\alpha$/Fe], at each wavelength. Hence, we construct new SSP models with different levels of $\alpha$-enrichment. Per [$\alpha$/Fe] value, we name this new set of models as a new "SSP family".  In the case of VCC0794, the most $\alpha$-enhanced dE in our sample with [$\alpha$/Fe] = 0.43 dex, the desired SSP family is constructed by linearly extrapolating the available SSP models beyond their maximum value of 0.4 dex. To fit each dE, we consider three $\alpha$ values (hence three SSP families): The derived [$\alpha$/Fe] and its corresponding upper and lower errors, as reported in Table \ref{Integrated values}. In Fig.\ref{STARLIGHTFIT_2019}, we show an example of the fit performed by STARLIGHT over the integrated spectrum of VCC2019 (the dark blue spectrum) using three SSP families with [$\alpha$/Fe] = [0.22, 0.26, 0.30] dex. The best-fitted solution of STARLIGHT (shown in red) and the fit residuals (shown in light purple) are also plotted in this figure. Please note that the residuals are shifted up by 0.3 for displaying purposes. 

To account for possible systematic errors, we first perform a single fit over the integrated spectrum of each dE. By randomly selecting a value from the residuals of this initial fit and adding that to the original flux, at each wavelength, we create 100 perturbed spectra and repeat the fitting on each of them. The results of our procedure are discussed in Section \ref{res:3}.


\section{Results}\label{Results}
\subsection{Integrated stellar properties}\label{Results:Integrated values}

As summarized in Table \ref{Integrated values}, our dEs span the age range of 2.10 to 10.50 Gyr, the metallicity range -0.77 $<$ [M/H] [dex]$<$ -0.37 and the [$\alpha$/Fe] interval between 0.14 and 0.43 dex. The youngest stellar populations, with an average age $=$ 2.10 Gyr and a mean [M/H] = $-$0.61 dex, are detected in VCC0170, while the oldest stellar populations are observed in VCC1896 and VCC2019. VCC1836 is the most metal$-$poor dE in our sample with [M/H] = $-$0.77 dex, while VCC1833 the most metal-rich with [M/H] = $-$0.37 dex. Additionally, VCC0794 is the most $\alpha$-enhanced member of our sample with [$\alpha$/Fe] = 0.43 dex. Given the stellar mass range of our sample, we find our derived [M/H] values for this sample to be in good agreement with the general  mass$-$metallicity scaling relation \citep[e.g.,][]{2005Gallazzi, 2008Panter,2014Delgado, 2017Sybilska}.

VCC0170, VCC0794, and VCC0990 have been investigated as part of the SMAKCED project by \cite{2014Toloba}. We find our measurements for VCC0170 and VCC0794 to be in good agreement with their results. In detail, \cite{2014Toloba} reported a light-weighted age of 2.0$\pm$0.4 Gyr (compared to 2.10$\pm$0.20 in this study) and [M/H] of -0.50$\pm$0.10 dex (compared to -0.61$\pm$0.04 dex derived here) for VCC0170. Similarly, they reported a light-weighted age of 7.8$\pm$1.8 Gyr (compared to 9.90$\pm$0.40  in this study) and [M/H] of -0.8$\pm$0.1 dex (compared to -0.73$\pm$0.02 dex obtained here) for VCC0794. However, we estimate an older age for VCC0990 (7.77$^{+3.59}_{-1.59}$ Gyr) than what was determined by \cite{2014Toloba} (4.0$\pm$1.2 Gyr). Nevertheless, this discrepancy is less than 2$\sigma$ and not significant.

\begin{figure*}
\includegraphics[scale=0.65]{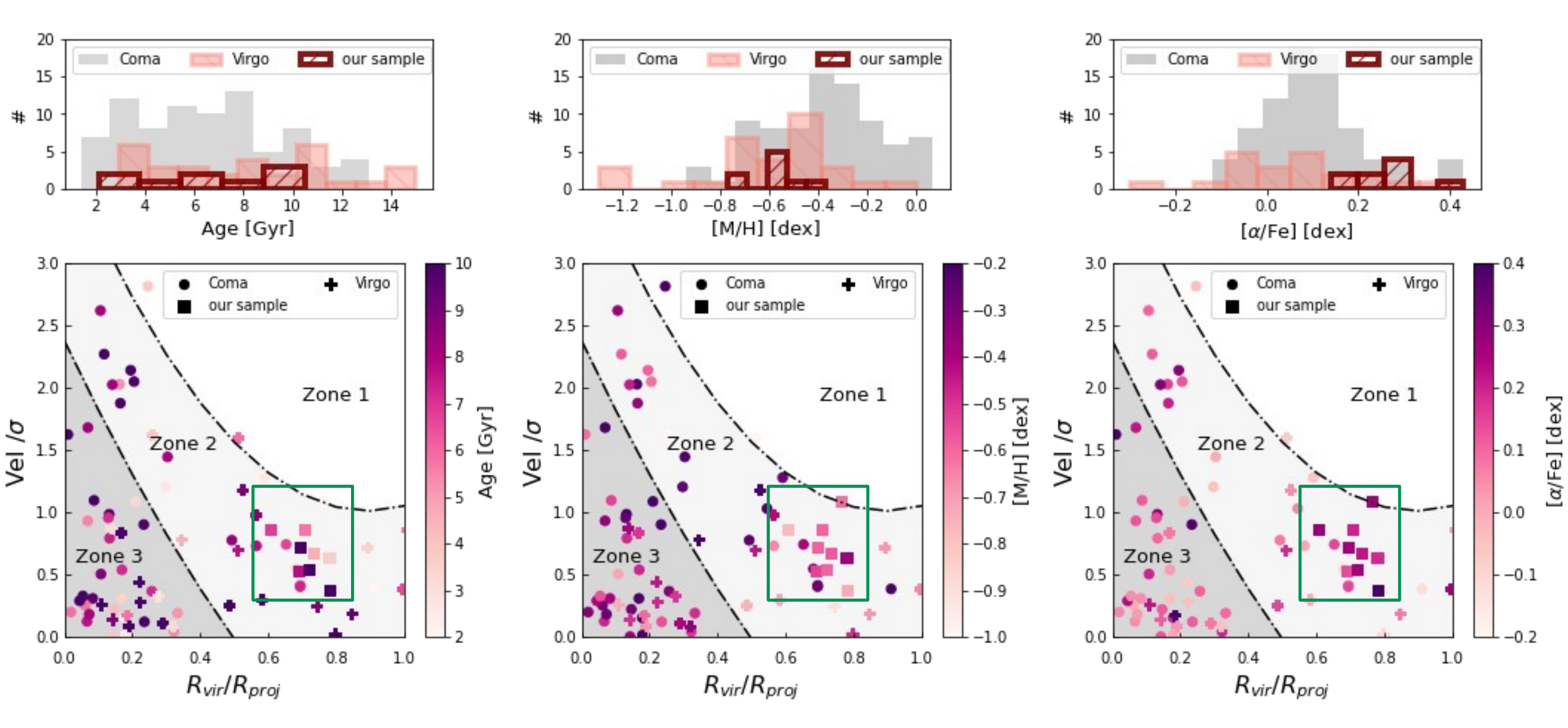}
\caption{\textit{Top row}: From left to right, the age, [M/H], and [$\alpha$/Fe] distributions of the Coma dEs (in gray), the Virgo dEs (in pink) and our sample of dEs (in brown) are presented. All dEs have -17$\geq$M$_{r}$ [mag]$>$-18. \textit{Bottom row}: From left to right, the projected phase-space distributions of the Virgo dEs (denoted with a plus symbol), Coma dEs (denoted with circle) and our sample of dEs (denoted with squares) colour-coded for age, [M/H] and [$\alpha$/Fe] in each panel, respectively. Our dEs are further marked with a green solid-line rectangle. Different zones of the projected phase-space diagram indicate different average infall times ($\protect\overline{\rm T}_{\rm inf}$) to the host halo, obtained from \protect\cite{2019Pasquali} and \protect\cite{2019Smith}. In detail: zone 1: $\protect\overline{\rm T}_{\rm inf}$ $<$ 3 Gyr; zone 2: 3 Gyr $<$ $\protect\overline{\rm T}_{\rm inf}$ $<$ 5 Gyr; zone 3: $\protect\overline{\rm T}_{\rm inf}$ $>$ 5 Gyr.}
\label{phase-space_stellarpop}
\end{figure*}

In Fig.\ref{phase-space_stellarpop} we compare the stellar population parameters of our sample of dEs with those of our comparison samples (see Section \ref{Data}). On the top row and from left to right, we present the age, [M/H] \footnote{To convert the [Fe/H] values reported in \cite{2009Smith} to [M/H], which we use in this study, the conversion of $[\rm M/H] \sim [\rm Fe/H] + \rm log(0.694f_{\alpha} + 0.306)$, is performed where $f_{\alpha} = 10^{[\rm \alpha/Fe]}$ \citep{2005Salaris}}, and [$\alpha$/Fe] distributions of the Coma dEs (in gray), the Virgo dEs (in pink) and our sample of dEs (in brown). In terms of light-weighted age distribution, our dEs are not significantly different from other dwarf ellipticals in the Virgo and Coma clusters. The middle upper panel of Fig.\ref{phase-space_stellarpop} shows that, despite their similarities in age, Virgo dEs (including our dEs) are more metal-poor than their counterparts in the Coma cluster (with a median [M/H] of -0.36$\pm$0.02 dex and -0.59$\pm$0.04 dex for the Coma and Virgo clusters, respectively). This can be either due to the lower number of investigated Virgo dEs within the given stellar mass range or the different characteristics of these two clusters (particularly their dynamical stage and halo size). Nonetheless, investigating the different characteristics of these two massive galaxy clusters is beyond the scope of this paper. 

Note that, remarkably, the distribution of [$\alpha$/Fe] in the right panel indicates that our dEs are more $\alpha$-enhanced (with a median [$\alpha$/Fe] = 0.26 $\pm$ 0.02 dex) than those in the Virgo and Coma clusters (with a median [$\alpha$/Fe] of 0.05$\pm$ 0.03 dex and 0.10$\pm$0.01 dex, respectively).

\begin{table*} \small%
\centering
\caption{\label{Phase-space-statistics} Median stellar population properties of our sample of dEs and the comparison samples in each infall zone}
\begin{tabular}{|p{2.3cm}|p{2.3cm}|p{1.5cm}|p{1.5cm}|p{1.5cm}| p{1.1cm}|}
\hline
Cluster & Median of & Zone 3 & Zone 2 \\
\hline
Coma & Age [Gyr]& 7.2 $\pm$ 0.54 & 7.55 $\pm$ 0.7 \\
     & [M/H] [dex]& $-$0.36 $\pm$ 0.03 & $-$0.31 $\pm$ 0.04 \\
     & [$\alpha$/Fe] [dex] & 0.12 $\pm$ 0.02 & 0.09 $\pm$ 0.03 \\
\hline
Virgo & Age [Gyr]& 9.57 $\pm$ 1.07 & 8.22 $\pm$ 0.8 \\
     & [M/H] [dex]& $-$0.49 $\pm$ 0.03 & $-$0.62 $\pm$ 0.06 \\
     & [$\alpha$/Fe] [dex]& 0.08 $\pm$ 0.04 & 0.05 $\pm$ 0.03 \\
\hline
Our sample & Age [Gyr]& --- & 6.8 $\pm$ 1.1  \\
     & [M/H] [dex]& --- & $-$0.59 $\pm$ 0.05  \\
     & [$\alpha$/Fe] [dex]& --- & 0.26 $\pm$ 0.02  \\
\hline
All data points$^{a}$ & Age [Gyr]& 7.40 $\pm$ 0.5 & 7.77 $\pm$ 0.5 \\
                & [M/H] [dex]& $-$0.40 $\pm$ 0.05 & $-$0.40 $\pm$ 0.03 \\
                & [$\alpha$/Fe] [dex]& 0.10 $\pm$ 0.02 & 0.08 $\pm$ 0.02 \\
\hline
\end{tabular} \\
The columns show: Name of the cluster, stellar population property for which we are reporting median values, median values in zone 3, and median values in zone 2. \\
a: Computed after excluding our sample of dEs. \\
\end{table*}

The projected phase-space diagram of a cluster, which combines cluster-centric velocity with cluster-centric radius, allows us to estimate
an average infall time for the cluster's galaxies, and hence how long they
have been exposed to the cluster's environmental effects
\citep{2017Rhee, 2019Pasquali,2019Smith}. To check for a possible correlation between the stellar population properties of dEs and their infall time,
we present the projected phase-space distribution of our sample along with the Virgo and Coma comparison samples in the bottom row of Fig.\ref{phase-space_stellarpop}. In each panel and from left to right, the data points are colour-coded based on their age, [M/H] and [$\alpha$/Fe]. Different symbols indicate members of different clusters. In each panel, three different zones are marked, indicating different average infall times ($\overline{\rm T}_{\rm inf}$), obtained from \cite{2019Pasquali} and \cite{2019Smith}. In this regard, our zone 3 corresponds to $\overline{\rm T}_{\rm inf}$ > 5 Gyr, zone 2 indicates 5 < $\overline{\rm T}_{\rm inf}$ < 3 Gyr and zone 1 indicates $\overline{\rm T}_{\rm inf}$ < 3 Gyr. Our dEs are denoted with coloured squares and are marked with a solid-line rectangle.
Detailed information on the statistics of each zone is presented in Table \ref{Phase-space-statistics} and is also discussed below.

\textbf{Median age - }According to this table, the median age of all dEs in zone 3 is 7.4 $\pm$ 0.5 Gyr, and in zone 2, after excluding our sample of dEs, 7.77 $\pm$ 0.5 Gyr. Our dEs with a median age of 6.8 $\pm$ 1.1 Gyr are consistent with the dEs in both zone 2 and 3 (within the 1-$\sigma$ level). However, despite their similar stellar mass and average infall time, they show a remarkable scatter in their ages, which might be related to pre-processing in their previous host group. We elaborate more on this in Section \ref{Discussion1}. 

\textbf{Median metallicity - }The median light-weighted [M/H] of all cluster dEs shows no trend with their average infall time. All dEs in zone 3 have a median [M/H] = -0.40 $\pm$ 0.05 dex, and -0.40 $\pm$ 0.03 dex in zone 2. Our dEs are more metal-poor than the dEs of zone 2 and 3, with a median [M/H] = -0.59 $\pm$ 0.05 dex (within the 2-$\sigma$ level). More specifically in zone 2, our dEs are more metal-rich than Virgo dEs (with median [M/H] = -0.62 $\pm$ 0.06), but metal-poorer than Coma dEs (with median [M/H] = -0.36 $\pm$ 0.05). It should be noted, however, that the statistics presented for the Virgo dEs are possibly affected by the small number of data points available for this cluster.

\textbf{Median [$\alpha$/Fe] - }
The median [$\alpha$/Fe] value of the cluster dEs, excluding our sample of dEs, is
$\sim$ 0.10 dex in both zones 2 and 3, and does not show any evident correlation with the average infall time. 
Our dEs have a median value of 0.26 $\pm$ 0.02 dex, thus
are more $\alpha$-enhanced than the Coma and Virgo comparison samples, at any given infall time. Additionally, our dEs fall within a narrow range of 0.14$<$[$\alpha$/Fe] [dex]$<$0.43. These values are comparable with what is reported for giant early-type galaxies \citep[e.g.,][]{2006Gallazzi,2014LaBarbera,2021Gallazzi} and thick disk stars of the Milky Way \citep[e.g.,][]{2003Bensby,2021Vincenzo}.

\subsection{On the [$\alpha$/Fe] ratio of our sample dEs}\label{result:alpha/Fe}

\begin{figure}
    \centering
    \includegraphics[scale=0.7]{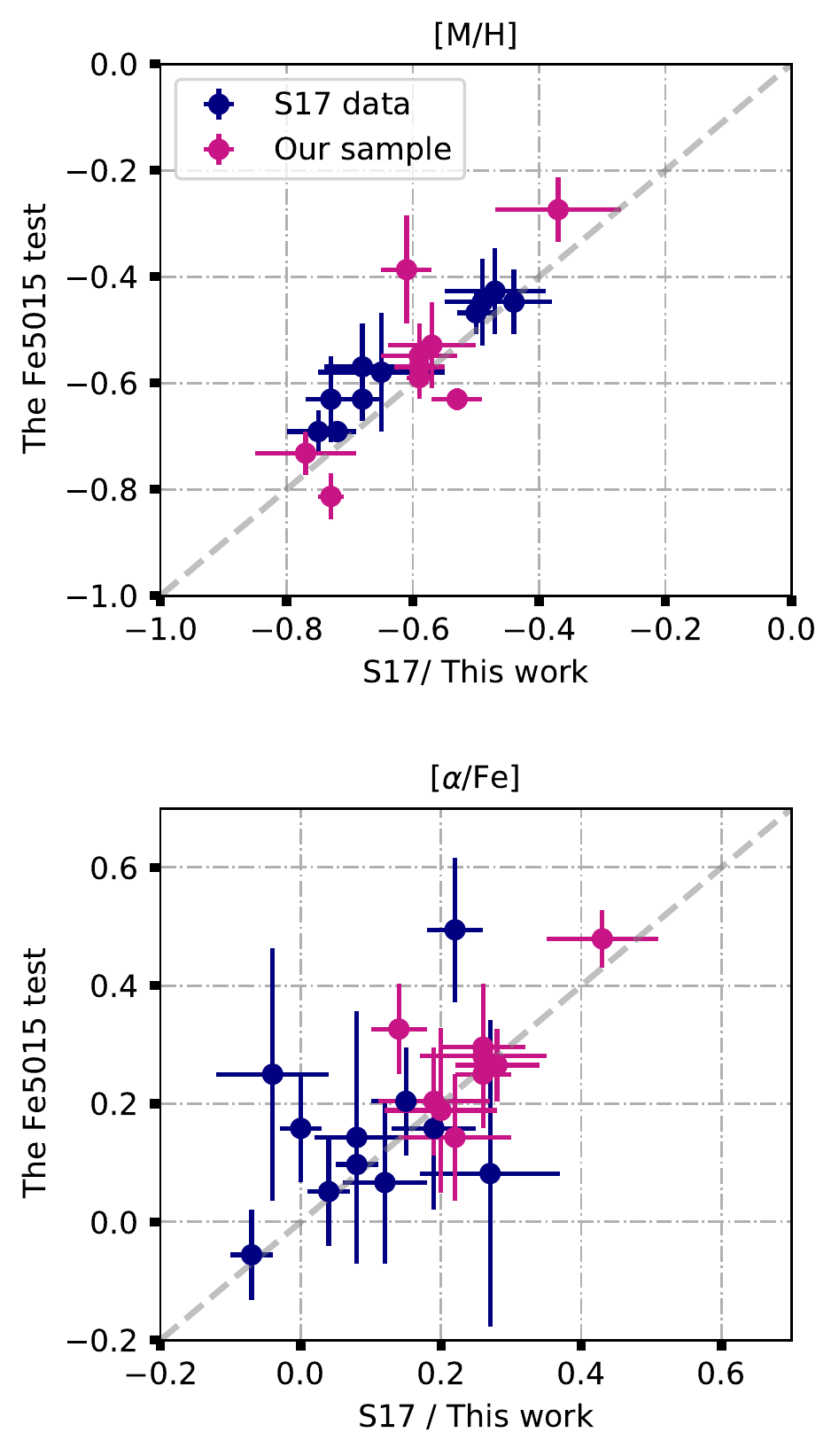}
    \caption{Metallicities and [$\alpha$/Fe] ratios obtained from Step 1 and 2 of our method, relying on the Fe5015 index, are shown on the y-axis of the top and bottom panels, respectively. On the x-axis of each panel, we compare the Fe5015 test results with the original estimates of \protect\cite{2017Sybilska} (S17, blue data points) and the values of Table \protect\ref{Integrated values} for our sample of dEs (pink data points). Error bars are computed in the same way as described in Section \protect\ref{Analysis}.}
	\label{Fe5015_test}
\end{figure}

Despite having a similar stellar mass, our dEs show larger [$\alpha$/Fe] values compared to their counterparts in the Coma and Virgo clusters. One possible reason for this difference can be the use of different Fe absorption lines to estimate the [$\alpha$/Fe] ratio. We estimate [$\alpha$/Fe] through the combination of the Mgb, Fe5270, and Fe5335 Lick indices, similarly to \cite{2009Smith} in their analysis for dEs in Coma. However, the [$\alpha$/Fe] ratios derived by \cite{2017Sybilska} are based 
only on the Mgb and Fe5015 indices. 

We therefore repeat our measurements using the same indices as in \cite{2017Sybilska}. To be consistent with their method, we derive age, [M/H], and [$\alpha$/Fe] only by going through the first two Steps described in Section \ref{Method: fitting}, and by using only the H$\beta_{0}$, Mgb, and Fe5015 indices. We apply this approach to our sample of dEs and \cite{2017Sybilska} sample. A one-to-one comparison of the derived [M/H] and [$\alpha$/Fe] values in this test with the original measurements is presented in the top and bottom panels of  Fig.\ref{Fe5015_test}, respectively.

On the y-axis of each panel in Fig.\ref{Fe5015_test}, we show the [M/H] ([$\alpha$/Fe]) values obtained only from Steps 1 and 2 of our method, relying on the Fe5015 index, for both our sample of dEs and \cite{2017Sybilska} sample. This set of values is labelled as “The Fe5015 test”. On the x-axis, we compare ``The Fe5015 test'' results with the original measurements of \cite{2017Sybilska} (S17, shown with blue data points). Pink data points show the comparison between ``The Fe5015 test'' results and our original estimates for our sample of dEs (obtained from Table \ref{Integrated values}). We find that, despite changing Lick index pairs in our fitting procedure, our dEs still show a distinct [$\alpha$/Fe] distribution, albeit with a slightly different range of values. Specifically, by fitting only over the Fe5015 absorption line, our sample of dEs falls within the range of 0.18$<$ [$\alpha$/Fe] [dex] $<$ 0.49, instead of 0.14$<$ [$\alpha$/Fe] [dex] $<$ 0.43 as obtained with our original method.

This comparison shows that choosing different Fe absorption features has negligible effects on the derived [$\alpha$/Fe] values of our dEs. Hence, the distinct [$\alpha$/Fe] distribution of our sample dEs, compared with \cite{2017Sybilska}'s sample, is not due to systematics. We compare the [$\alpha$/Fe] values of our sample of dEs (derived in this section) with those of \cite{2017Sybilska}, using the K-S and the Anderson-Darling tests. Both tests indicate that the difference between the [$\alpha$/Fe] distributions of the two samples is significant (with p-values = 0.03 and 0.009, respectively).

\setlength{\tabcolsep}{10pt}
\begin{table*}
\caption{\label{Starlight_age_metal} Integrated light-/mass-weighted stellar population properties of our sample dEs as derived by STARLIGHT}
\centering
\small
\begin{tabular}{c c c c c c}
\hline
Object  & M$_{\rm B}$  &$<$Age$>$$_{\rm L}$ & $<$[M/H]$>$$_{\rm L}$ & $<$Age$>$$_{\rm M}$ & $<$[M/H]$>$$_{\rm M}$ \\
    & [mag] & [Gyr] & [dex] & [Gyr] & [dex] \\ 
\hline
\hline
VCC0170& -16.63 &2.09$^{+0.20}_{-0.24}$ & $-$0.54$^{+0.04}_{-0.04}$& 2.78 $^{+0.21}_{-0.19}$& $-$0.61 $^{+0.12}_{-0.10}$ \\[0.2cm]
VCC0407& -16.38 & 4.11$^{+0.35}_{-0.42}$ & $-$0.50$^{+0.03}_{-0.03}$& 4.70 $^{+0.31}_{-0.34}$ & $-$0.56 $^{+0.13}_{-0.11}$
\\[0.2cm]
VCC0608& -16.59 & 3.16$^{+0.27}_{-0.66}$ & $-$0.44$^{+0.04}_{-0.04}$& 4.58 $^{+0.20}_{-0.34}$ & $-$0.48 $^{+0.14}_{-0.11}$
 \\[0.2cm]
VCC0794& -16.30 & 9.54$^{+0.58}_{-0.52}$ & $-$0.69$^{+0.05}_{-0.05}$& 9.91 $^{+0.83}_{-0.78}$ & $-$0.62 $^{+0.14}_{-0.10}$
\\[0.2cm]
VCC0990& -16.44 & 4.37$^{+0.43}_{-0.50}$ & $-$0.38$^{+0.04}_{-0.04}$& 5.30 $^{+0.40}_{-0.38}$ & $-$0.37 $^{+0.09}_{-0.11}$
\\[0.2cm]
VCC1833& -16.45 & 5.16$^{+0.20}_{-0.82}$ & $-$0.28$^{+0.08}_{-0.08}$& 7.17 $^{+0.36}_{-0.25}$ & $-$0.27 $^{+0.09}_{-0.09}$\\[0.2cm]
VCC1836& -16.46 & 3.98$^{+0.45}_{-0.38}$ & $-$0.66$^{+0.05}_{-0.05}$& 4.56 $^{+0.33}_{-0.28}$ & $-$0.72 $^{+0.20}_{-0.16}$\\[0.2cm]
VCC1896& -16.05 & 8.60$^{+0.43}_{-0.57}$ & $-$0.49$^{+0.04}_{-0.03}$& 9.06 $^{+0.76}_{-0.70}$& $-$0.43 $^{+0.18}_{-0.10}$\\[0.2cm]
VCC2019& -16.66 & 6.51$^{+0.60}_{-0.65}$ & $-$0.46$^{+0.05}_{-0.04}$& 6.81 $^{+0.49}_{-0.41}$& $-$0.45 $^{+0.12}_{-0.08}$\\[0.2cm]
\hline
\end{tabular}\\
The columns show: name of the target, total R-band magnitude, light-weighted age, light-weighted metallicity, mass-weighted age, and mass-weighted metallicity. M$_{\rm B}$ is derived from the observed M$_{\rm r}$ 
using the \textit{B-r} colour computed by MILES SSPs with the same age and metallicity as estimated for each dE. \\
\end{table*}

\subsection{Star formation histories of our sample dEs}\label{res:3}

As explained in Section \ref{method:STARLIGHT}, we perform the full spectrum fitting using STARLIGHT, for 100 MC realizations over each dE's averaged spectrum. We derive light-weighted values of age and [M/H] from each MC run, following the definitions below: 

\begin{equation}
    <\rm Age>_{L} = \sum_{t,Z\rvert\alpha} \epsilon_{t,Z\rvert\alpha} \rm Age
\end{equation} 
\begin{equation}
    <\rm [M/H]>_{L} = \sum_{t,Z\rvert\alpha} \epsilon_{t,Z\rvert\alpha} \rm [M/H]
\end{equation}

\noindent where \textit{t, Z} and $\alpha$ denote age, metallicity and [$\alpha$/Fe] ratio of a given SSP model, respectively, and $\epsilon_{t,Z\rvert\alpha}$ is the light-weighted stellar population vector. The latter indicates the flux contribution of each SSP model to the best-fit spectrum. Here the summation is over the entire base models, at fixed $\alpha$. To compute the mass-weighted age and metallicity, we use the same equations but
with $\mu_{t,Z\rvert\alpha}$, which is the mass-weighted stellar population vector. This vector is computed through multiplying $\epsilon_{t,Z\rvert\alpha}$ by the mass-to-light ratio of each SSP (computed by STARLIGHT for each fit). In Table \ref{Starlight_age_metal} we report the light- and mass-weighted ages and metallicities averaged over the results of the 100 MC runs. Their upper and lower errors are the 84th and 16th percentiles of their corresponding distributions. In Appendix \ref{app:2} we compare the values of light-weighted age and [M/H] that were obtained through full spectrum fitting with those derived by our method described in Section \ref{Method: fitting}. The comparison shows that the differences in age 
and [M/H] results between our index fitting method and STARLIGHT are typically significant at a 2$\sigma$ level. STARLIGHT reports, on average, younger ages and higher [M/H]. The largest discrepancy concerns the age of VCC2019, for which STARLIGHT estimates a value of 6.5 Gyr (compared to $\sim$10.5 Gyr from our index fitting method). In Appendix \ref{app:3} we show that the presence of a young nuclear star cluster in the core of VCC2019 leverages the contribution of younger SSPs in the STARLIGHT final best fit for this dE. The central nuclear star cluster hosts a distinct, younger stellar population compared to the rest of VCC2019 and noticeably contributes to its integrated light \citep[see][]{2010Paudel,2021Fahrion}.

\begin{figure*}
    \centering
	\includegraphics[scale=0.55]{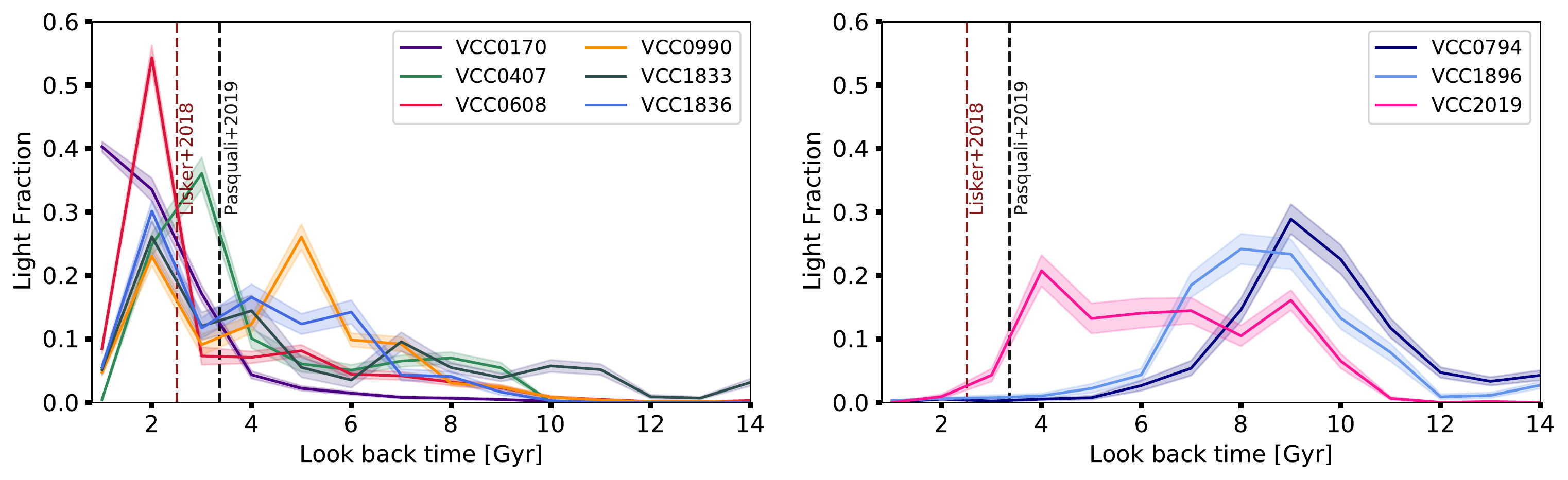}
	\includegraphics[scale=0.55]{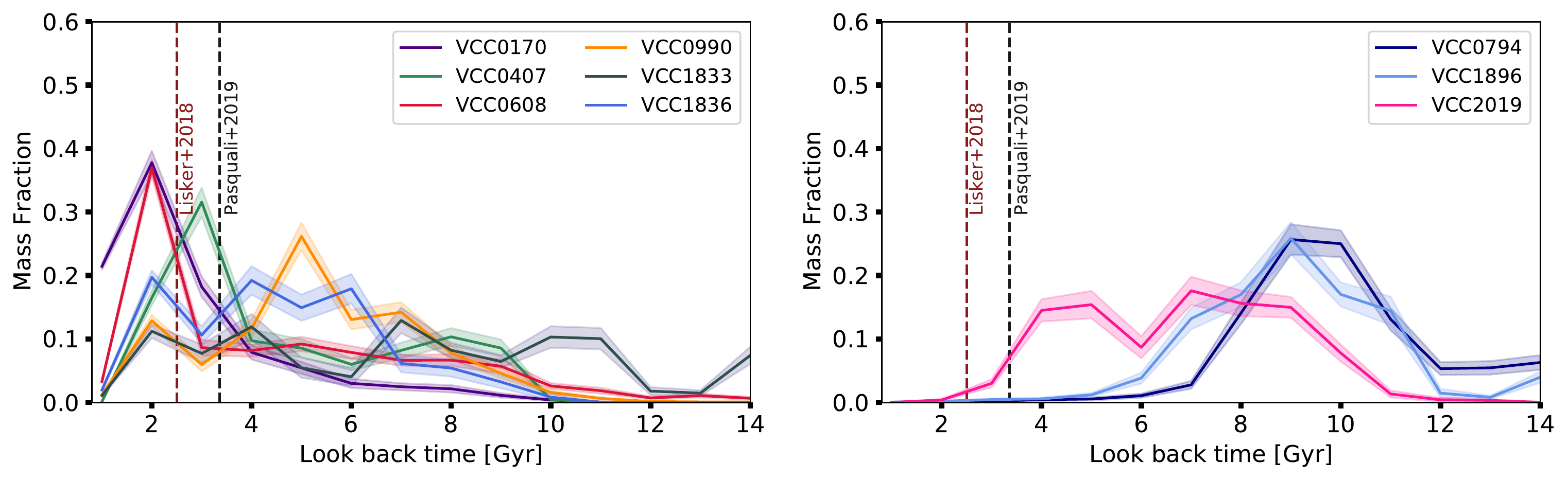}
	\caption{\textit{Top row:} The stellar light fraction ($\epsilon_{t,Z\rvert\alpha}$) of our sample of dEs as a function of time. The profiles of those dEs that exhibit a prominent peak after their accretion onto the Virgo cluster are presented in the left-hand panel, and the distributions of those dEs without a similar peak are shown in the right-hand panel. \textit{Bottom row:} Same as the top row but for the mass weighted fractions ($\mu_{t,Z\rvert\alpha}$). The red and black dashed vertical lines indicate the accretion time of our sampled dEs onto Virgo as predicted by \protect\cite{Lisker2018} and \protect\cite{2019Pasquali}, respectively.} 
	\label{ligh-mass-fractions}
\end{figure*}

 \begin{figure}
     \centering
 	\includegraphics[scale=0.5]{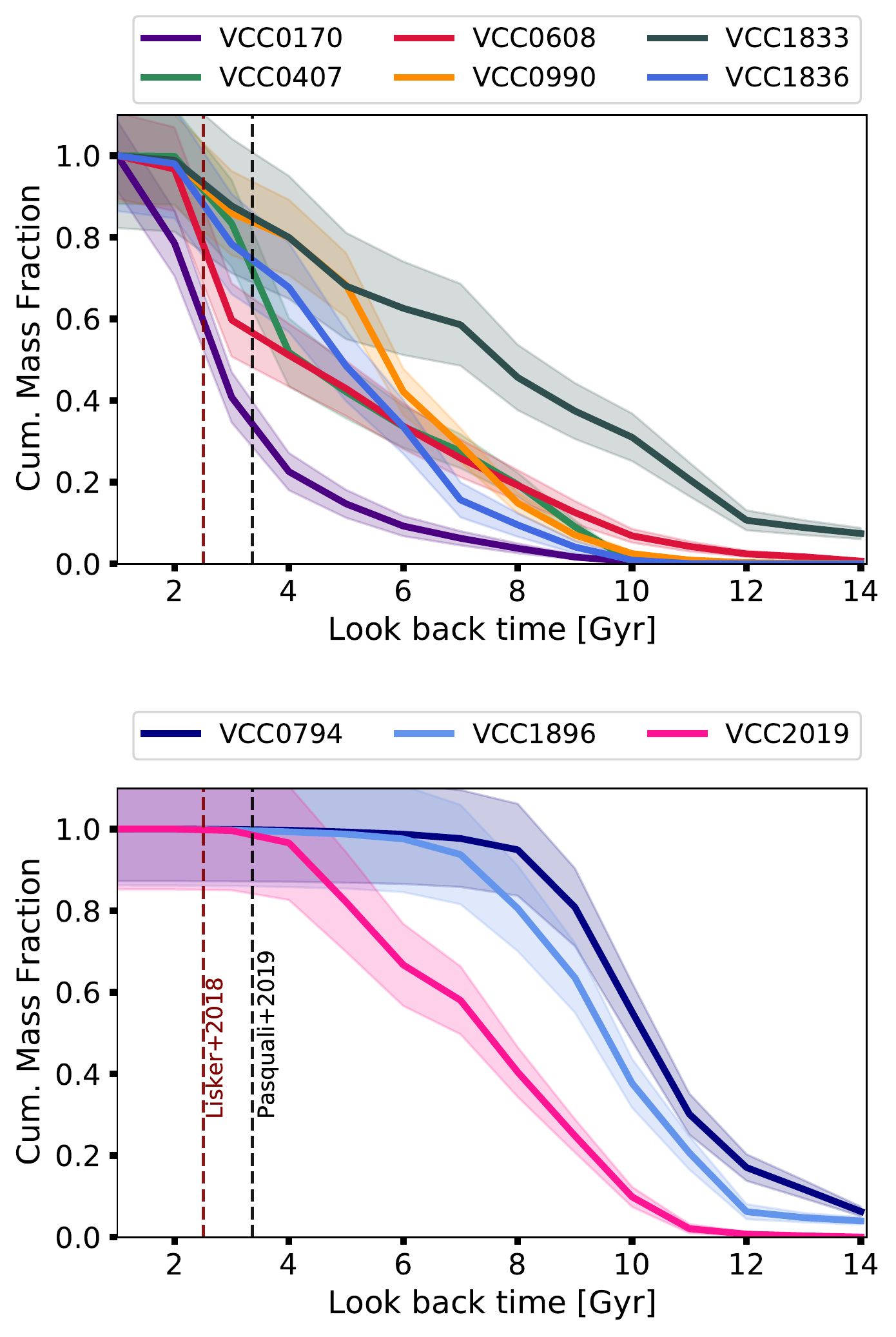}
 	\caption{The cumulative mass distribution as a function of look back time (in Gyr) plotted for our sample dEs with (top panel) and without (bottom panel) a recent episode of star formation. Here, galaxies are divided into two sub-samples following Fig. \protect\ref{ligh-mass-fractions}. The red and black dashed vertical lines indicate the accretion time of our sample of dEs onto Virgo as estimated by \protect\cite{Lisker2018} and \protect\cite{2019Pasquali}, respectively. } 
 	\label{cumulative_prof}
 \end{figure}
 
In the top row of Fig.\ref{ligh-mass-fractions} we present the stellar light fraction ($\epsilon_{t,Z\rvert\alpha}$) as a function of look-back time for each dE in our sample. Each distribution is constructed by summing $\epsilon_{t,Z\rvert\alpha}$ of the base models of the same age but different metallicity and [$\alpha$/Fe] ratio.
These light distributions are sensitive to the contribution of young stars to the total observed light. Thus, they trace the more recently formed stars in a given dE. Here we divide galaxies into two sub-samples based on the presence or absence of a peak in their light fraction profiles at $\sim$ 2-3 Gyr. In the lower panels of this figure, we present the stellar mass fraction ($\mu_{t,Z\rvert\alpha}$) as a function of look back time, which is constructed by summing $\mu_{t,Z\rvert\alpha}$ of the base models of the same age but different metallicity and [$\alpha$/Fe]. For each dE, the mass-weighted distribution follows a similar trend as the light-weighted one. In all four panels of Fig. 5, two dashed vertical lines indicate two estimates for the average infall time ($\overline{\rm T}_{\rm inf}$) of our sample dEs, based on the results of \cite{Lisker2018} (the red line) and \cite{2019Pasquali} (the black line).

The six dEs in the left-hand panels of Fig.\ref{ligh-mass-fractions} seem to have experienced a rather prolonged star formation activity at low rate, which has later undergone a sudden boost during or after their accretion onto Virgo.
On the contrary, two of the dEs in the right-hand panels of Fig.\ref{ligh-mass-fractions} (namely VCC0794 and VCC1896) exhibit a shorter star formation activity that culminated about 8 - 10 Gyrs ago. VCC2019 also follows a similar trend, after excluding its central nuclear star cluster from its integrated spectrum (see Appendix \ref{app:3}). 

In the top panel of Fig.\ref{cumulative_prof} we plot the cumulative mass fraction of these six dEs as a function of look-back time. Our results indicate that this sub-population has experienced distinguishable, yet diverse (ranging from 5 to more than 70 percent), mass growth since the accretion event (i.e., between 4 to 1 Gyr ago). According to Fig.\ref{cumulative_prof} VCC0170, where B20 detected ongoing star formation, has experienced the most drastic mass growth ( $\sim$ 70 percent) since $\sim$ 4 Gyr ago. 
Our results for these six dEs are in general agreement with the "delayed-then-rapid" quenching scenario of \cite{2013Wetzel} (see Section \ref{Discussion}), in which galaxies are predicted to grow in mass by up to 50 percent after their accretion onto clusters.
Regarding the remaining three dEs in our sample, their distributions of cumulative mass fraction in Fig.\ref{cumulative_prof} indicate that they formed their current total stellar mass more than 4 Gyr ago, before their accretion onto Virgo and possibly in their 
previous host group.

In Section \ref{Results:Integrated values}, we show that VCC0794 is the most $\alpha$-enhanced member of our sample. The spectral indices of this dE point to old ($\sim$ 10 Gyr) and metal-poor ($\sim$ -0.73 dex) stellar populations (Table \ref{Integrated values}). In the top right panel of Fig.\ref{ligh-mass-fractions}, this particular dE shows a peak in its stellar light fraction profile at the age of 9 Gyr, which is consistent with our previous results and endorses the fact that this dE has been accreted onto Virgo already old and quenched. The profile of this particular dE indicates that the bulk of its stellar population formed in a rather short time interval ($\sim$ 4 Gyr), which can explain its high value of [$\alpha$/Fe]. 

Two dEs of our sample, namely VCC0990 and VCC2019, show double peaks in their stellar light and mass fraction profiles. According to our results in Appendix \ref{app:3}, the peak at younger ages (i.e., at $\sim$ 2 Gyr for VCC0990 and $\sim$ 4 Gyr for VCC2019) is mostly due to their
nuclear star cluster, while the second peak at older ages appears to be related to star formation in their main body. We note that the exclusion of the nuclear star cluster from the galaxy's integrated spectrum (Fig.\ref{NSC_correction1}) significantly reduces the amplitude of the peak at younger ages in the star formation history of both galaxies. It thus appears that VCC2019 was accreted onto Virgo with its main body already pre-processed and quenched
\citep[also in][]{2021Fahrion}. A detailed investigation on the spatial distribution of stellar population properties of these two dEs (along with the rest of our sample) will be presented in a forthcoming paper. 

Our results, constructed from 100 MC iterations, present only 100 possible fitting solutions for the spectral energy distribution of each dE. In order to have a realistic interpretation of these results, however, we need to also take inevitable sources of uncertainty into account. Our full spectrum fitting, performed with STARLIGHT, is based on a limited number of SSP models, which may not be enough to break 
the well-known “age-metallicity degeneracy”, especially for 
ages older than 1 Gyr (Shen \& Yin 2020). Yet, \cite{2011MNRAS.415..709S} showed that the age-metallicity degeneracy is less prominent in the full spectrum fitting technique compared to the Lick indices method. This degeneracy may affect the accurate age dating of the stellar light fraction peaks in Fig.\ref{ligh-mass-fractions} (possibly shifting them by 1 to 2 Gyr), but given the uncertainties in the accretion time of our sample dEs, we believe that this does not significantly affect
our results and conclusions.


\section{Discussion}\label{Discussion}

Due to their shallow potential well, low-mass galaxies are quite susceptible to 
environmental processes, such as ram pressure stripping \citep{1972Gunn}, 
strangulation \citep{1980Larson}, and harassment \citep{1996Moore,1998Moore}, 
which can alter their morphology, star formation activity, and kinematics 
on different timescales \citep[e.g.,][]{2006Boselli,2021Boselli}. 
For instance, low-mass satellites grow older and metal-richer as the mass of their host halos increases, whilst their star formation rate declines and their [$\alpha$/Fe] abundance ratio remains unchanged \citep{2010Pasquali,2019Pasquali,2020Coenda,2020Tiwari,2021Gallazzi,2021Trussler}. In the projected phase-space diagram, cluster low-mass galaxies with larger infall time (i.e., longer exposures to the cluster environment) are observed to be dominated by older, metal-richer and more [$\alpha$/Fe] enhanced stellar populations, compared to their recently accreted counterparts \citep{2019Pasquali,2021Gallazzi}.

The aforementioned trends are mainly explained within the so-called “delayed-then-rapid” scenario \citep[e.g.,][]{2013Wetzel,2014Wheeler,2016Oman,2019Maier,2019Maier1,2020Rhee}. In this context, the star-formation rate of newly accreted low-mass galaxies remains unaffected for several Gyr after their first infall. During this “delayed phase”, low-mass star-forming galaxies are predicted to experience noticeable (up to 50 percent) mass growth \citep{2013Wetzel}. During the subsequent “rapid phase” they are predicted to reduce their star formation rate with an e-folding time of $<$ 0.8 Gyr \citep{2013Wetzel,2020Rhee}. The rapid phase is possibly driven by ram pressure stripping, which is more efficient in the central regions of massive galaxy groups and clusters. Furthermore, ram pressure can compress the gas in the inner galaxy disk \citep[e.g.,][]{2021Boselli} and, therefore, trigger a new episode of star formation or enhance a pre-existing one \citep{1998Fujita,1999Fujita,2001Vollmer}. It should be noted that the quenching time-scales predicted by this scenario are independent of host halo mass, and they can not reproduce the observed increase of the quiescent fraction of low-mass galaxies with halo mass. The scenario thus requires that up to 50 percent of the quenched low-mass galaxies in present-day host halos to be accreted already quenched prior infall, presumably in lower mass halos. 

In what follows, we use the stellar population properties and star formation histories of our sample of dEs to discuss their pre-processing within their previous group environment and their on-going processing within Virgo, their present-day host halo.

\subsection{The high [$\alpha$/Fe] ratio of dEs and their infall time}\label{Discussion1}
The [$\alpha$/Fe] abundance ratio is typically interpreted as an indicator of the duration of star formation in a given galaxy prior to the onset of significant contributions from SNe of Type Ia, in the sense that rapid or early quenching of star formation, can result in a higher [$\alpha$/Fe] ratio \citep[e.g.,][]{1998Balogh,1999Thomas,2011deLaRosa, 2015Santos}. Since dEs are more prone to environmental effects, one can expect a correlation between their chemical abundance ratio and their host environment. Yet, results in this regard seem contradictory. For instance, \cite{2008Michielsen} did not detect any correlation between the [$\alpha$/Fe] ratios and local density of very bright Virgo dEs (i.e., M$_{\rm r}$ <-15.5 mag) \citep[see also][]{2005Thomas,2010Paudel}. On the other hand, \cite{2009Smith} showed that the [$\alpha$/Fe] ratio of dEs in the Coma cluster decreases toward larger cluster-centric distances, from [$\alpha$/Fe] $\sim$ 0.15 dex in the cluster center to [$\alpha$/Fe]$\sim$ 0.0 dex in the cluster outskirts. A similar trend was also observed for Virgo dEs \citep{2016Liu} and low-mass galaxies in Abell 496 \citep{2008Chilingarian}, albeit with high uncertainty \citep[see also][]{2008Penny}. 

We note that the [$\alpha$/Fe] ratios obtained here for our sample of dEs contrast with the findings mentioned above as well as with the [$\alpha$/Fe]-$\overline{\rm T}_{\rm inf}$  correlation found by \cite{2019Pasquali} \citep[see also][]{2021Gallazzi}. In spite of their large cluster-centric distance ( $\sim$ 1.5 Mpc from M87) and small $\overline{\rm T}_{\rm inf}$, our dEs show high (not low) [$\alpha$/Fe] ratios (i.e., $>$ 0.2 dex), which are more similar to those of ancient infallers (with $\overline{\rm T}_{\rm inf}$ > 5 Gyr \cite{2019Pasquali}) or even to those of massive ETGs \citep[e.g.,][]{2006Gallazzi,2014LaBarbera,2021Gallazzi}. We find our sample of dEs to be more $\alpha$-enhanced (at a 8$\sigma$ level) than their counterparts in the Virgo and Coma clusters, at similar or even larger $\overline{\rm T}_{\rm inf}$ . Moreover, our dEs are $\alpha$-enhanced irrespectively of their luminosity-weighted ages. Such high [$\alpha$/Fe] ratios might be explained if our dEs had been affected by environmental processes not only during, but also before their accretion onto Virgo. We elaborate more on this in Section \ref{Discussion2}.

\subsection{Pre-processing in the group vs. early processing in Virgo}\label{Discussion2}

As shown by the star-formation histories (SFHs) of our sample of dEs (Section \ref{res:3}), six of these galaxies have undergone a  recent  phase of  star formation, during  or right after their  accretion onto  the  Virgo cluster (hereafter, rSF-dEs), while the  remaining three dEs show no signs of "recent" star formation,  having  ceased to form  stars long  before their accretion onto Virgo (hereafter, pSF-dEs).

The high [$\alpha$/Fe] of rSF- and pSF-dEs could be explained as follows:
\begin{description}

\item[ {\it rSF-dEs.} ] These galaxies experienced  distinguishable, yet diverse (ranging from 5 to more than 70 percent), mass growth within a short period  of time during/after  the accretion event (i.e., between 4 to 1 Gyr ago). The short duration of recent star formation in the rSF-dEs is also consistent  with these galaxies having similar (not higher) metallicities than the rest of our sample. Hence, at least some of the rSF-dEs could have high light-weighted [$\alpha$/Fe], because of the short star-formation event that they experienced after accretion onto Virgo.

\item[ {\it pSF-dEs.} ] These galaxies do not show any peak in their recent SFH, and completed the formation of their present-day stellar  mass more than 4 Gyr ago
(see right-hand panel of Fig.~\ref{ligh-mass-fractions}). In particular, VCC0794 formed most of its stellar mass $\sim$ 5 Gyr before accretion onto Virgo, within a time-scale of $\sim$ 2 Gyr~\footnote{This can  be roughly estimated, by eye, as the time since the star formation ceased after its peak value.}, in agreement with its high [$\alpha$/Fe] ($\sim 0.43$ dex) and low metallicity ($\sim$ -0.73 dex; see Table~\ref{Integrated values}). We thus suggest that the high [$\alpha$/Fe] ratio of the pSF-dEs might be the result of 
pre-processing in  their host group prior infall into Virgo.
\end{description}

However, as  the  sub-samples of rSF- and pSF-dEs exhibit very similar metallicities  and  [$\alpha$/Fe]  ratios (see Table \ref{Integrated values}), their [$\alpha$/Fe] values might have a common origin. Indeed, metallic lines are far shallower in young, relative to old, stellar populations  \citep{2009Walcher,2013Conroy}, and  therefore our [M/H] and [$\alpha$/Fe] estimates might be reflecting, for all galaxies, the properties of the old component, which was pre-processed in their host group prior accretion onto Virgo,
as discussed above for the sample of pSF-dEs. In order to distinguish between the two scenarios (early effect of Virgo vs. pre-processing in the previous host group) one would need to derive the chemical enrichment history of our sample dEs which, however, requires a larger spectral range (more extended to the blue) than provided by MUSE.

Regarding the origin of the recent star formation undergone by the rSF-dEs, we speculate that
the ram pressure exerted by the Virgo intra-cluster medium on these galaxies might be
responsible for it. In fact, by compressing the cold gas in the inner disc of galaxies during their early accretion, ram pressure can increase the central gas surface density by up to a factor of 1.5, and lead to a temporary enhancement of the star-formation rate by up to a factor of 2 in the central regions of the infalling galaxies \citep[][]{2001Vollmer,2020Steyrleithner}.
The presence of nebular emission lines powered by ongoing star formation in the spectra of VCC0170 might be a signpost of such an effect. On the other hand, the non-detection of ionized gas in the MUSE spectra of the remaining rSF-dEs might indicate that these galaxies had a lower gas reservoirs at infall,  
and/or consumed it faster. An alternative to our interpretation is that these recent episodes of star formation might be the result of the intrinsic star formation activity of these galaxies.

Since we interpret the high [$\alpha$/Fe] of the pSF-dEs in 
our sample as an indication of their pre-processed nature, 
the smaller [$\alpha$/Fe] values of the Virgo and Coma comparison samples in Fig. \ref{phase-space_stellarpop} may suggest
that pre-processing is not that important for the overall population of cluster dEs, despite predictions of a significantly large fraction of pre-processed galaxies in clusters \citep[e.g.,][]{2009McGee,2013Wetzel}.
However, the level of pre-processing depends on the gas content of a galaxy, the mass of its previous host halo as well as on its infall time onto such halo. Hence, determining the number of dEs with similar level of pre-processing as in our sample is very challenging, and is hampered by the limited size of our comparison samples, particularly in the Virgo cluster. Based on our results, one should not conclude that pre-processing is a rare condition in present-day clusters.
We also note that 
the  high [$\alpha$/Fe] of the rSF-dEs in our sample is likely due to a rather short and recent phase of star formation, taking place less than $\sim$ 4 Gyr ago. It is thus a recent event that cannot be traced in dEs with larger $\overline{\rm T}_{\rm inf}$ (i.e., zone 3) or those pre-processed recent infallers that are already gas deficient before their accretion onto the cluster (i.e., zone 2 and 1).

\subsection{Linking kinematics and stellar age of dEs} \label{Discussion3}

\begin{figure}
     \centering
 	\includegraphics[scale=0.8]{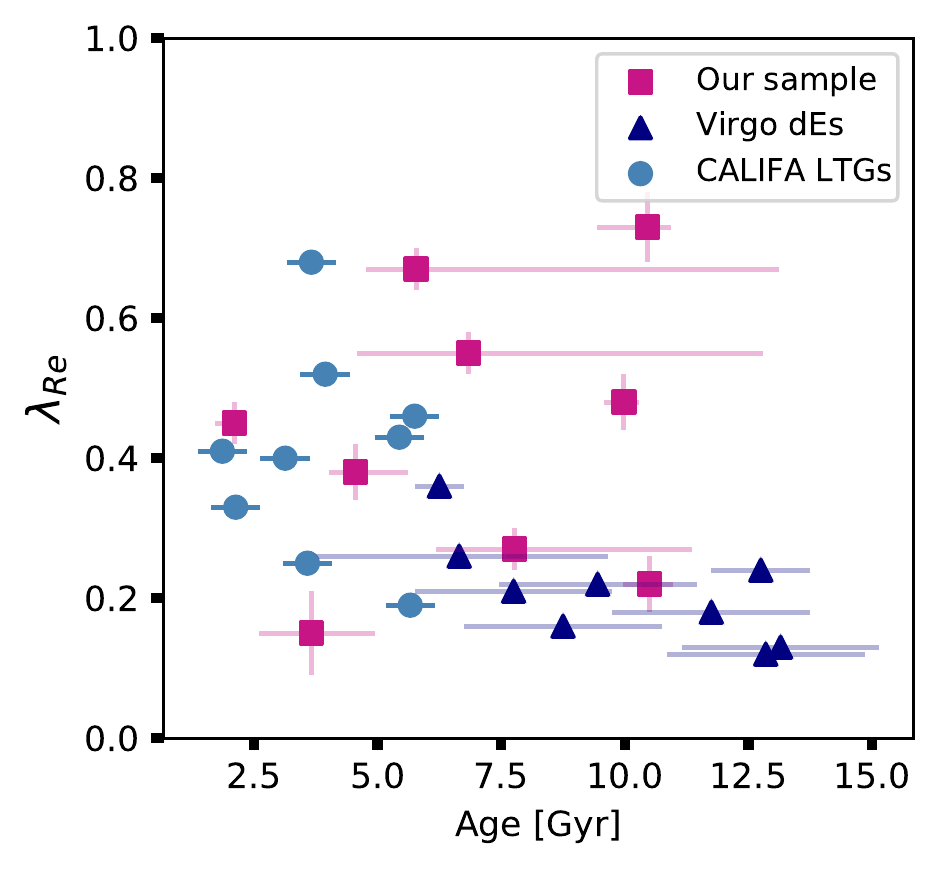}
 	\caption{This plot shows the specific angular momentum at 1R$_{e}$ ($\lambda_{\rm Re}$) vs. mean light-weighted age within 1R$_{e}$ of our sample of dEs and two comparison samples. Light blue circles indicate low-mass late-type field galaxies selected from the CALIFA sample. The $\lambda_{\rm Re}$ and age values are taken from \protect\cite{2019FalconBarroso} and \protect\cite{2015Delgado}, respectively. Dark blue triangles denote Virgo dEs with $\overline{\rm T}_{\rm inf}$ $>$ 3 Gyr. The $\lambda_{\rm Re}$ and age values for this sample are taken from \protect\cite{2014Rys} and \protect\cite{2017Sybilska}, respectively. Pink squares indicate members of the nine dEs investigated in this study. The $\lambda_{\rm Re}$ and age values are taken from B20 and Table \protect\ref{Integrated values}, respectively. } 
 	\label{Age_lambda}
 \end{figure}

In B20 we showed that the specific angular momentum ($\lambda_{R}$) profiles of our sample of dEs vary between those of low-mass, star-forming galaxies in the field and
those of equally-massive Virgo dEs with larger infall times.
We interpreted the scatter in $\lambda_{R}$ among our sample of dEs as possibly due to 
pre-processing in their host group prior to infall into Virgo, since the Virgo cluster can modify the kinematics of its satellites only after several pericenter passages \citep[i.e., several Gyr, ][] {2006Boselli}, while our dEs were accreted less than 3 Gyr ago. 

In Fig.\ref{Age_lambda}, we plot $\lambda_{\rm Re}$, the specific angular momentum measured at 1R$_{e}$, as a function of mean light-weighted age (within 1R$_{e}$) for Virgo dEs (in dark blue, \cite{2014Rys}; \cite{2017Sybilska}) and low-mass, late-type galaxies
in the field from the CALIFA survey (in light blue, \cite{2019FalconBarroso}; \cite{2015Delgado}). A detailed description of the field comparison sample is presented in B20. Here we note that the light-weighted ages of the Virgo dEs sample were derived using the Lick indices, while the light-weighted ages of the CALIFA sample were obtained using the full spectrum fitting technique with STARLIGHT. For our sample we use the light-weighted ages computed with our Lick indices fitting routine.
The Virgo dEs shown in this plot have larger infall times ($\overline{\rm T}_{\rm inf}$ > 3 Gyr) than our sample of dEs, while the CALIFA late-type galaxies can be considered to have $\overline{\rm T}_{\rm inf} =$ 0.

We see that these two comparison samples define a trend of decreasing $\lambda_{\rm Re}$ with increasing light-weighted age. As star formation gradually fades due to gas consumption and/or gas loss triggered by the environment, galaxy kinematics progressively changes from being dominated by rotation to being pressure-supported. The $\lambda_{\rm Re}$ - age relation of 
Fig.\ref{Age_lambda} is consistent with that derived by \cite{2019FalconBarroso} over a larger stellar mass range, and 
is also predicted by semi-analytic models of galaxy formation and evolution, according to which gas rich galaxies (with M$_{\star}$/ M$_{\odot}$ $>$ 10$^{9}$) have larger $\lambda_{\rm Re}$ \citep{2018Zoldan}.

Our dEs (denoted with pink squares) generally follow the relation defined by our two comparison samples, except for two
galaxies that are several $\sigma$ away from it, being too old ($\sim 10 \rm \, Gyr$)
for their high $\lambda_{\rm Re}$.
 
Under the assumption that their accretion onto Virgo is too recent for the Virgo cluster to have modified their kinematics, we qualitatively interpret the distribution of our sample dEs in the 
$\lambda_{\rm Re}$-age plane as follows. The older ($\gtrsim$ 5 Gyr) and lower $\lambda_{\rm Re}$ ($\lesssim$ 0.4) dEs might have fallen into their previous host
group earlier (and/or have been less gas-rich) than the other dEs in our sample.
In this way their group environment had enough time to quench their
star formation activity (or quenched it faster), and to transform their kinematics to what we have measured. Conversely, the younger dEs in our sample ($\lesssim$ 5 Gyr) might have been accreted at later times (and/or have been more gas-rich), so that their previous host group was able to transform them to a lesser extent. Their different $\lambda_{\rm Re}$ values might reflect their different star-formation rates, in the sense that a higher star-formation rate might yield both a young light-weighted age and a high $\lambda_{\rm Re}$, while a low star-formation rate might not be sufficient to bear a high $\lambda_{\rm Re}$. Additionally, the accretion onto Virgo and the consequent ram pressure of the cluster might have contributed to keep
$\lambda_{\rm Re}$ high by temporarily increasing the star-formation rate of the younger dEs in our sample (see Sect. \ref{Discussion2}).

The two outliers in our sample in Fig.\ref{Age_lambda} are VCC2019 and
VCC0794. It should be noted that VCC2019 shows a star formation history similar to that of VCC0794, after excluding its central nuclear star cluster (Fig.\ref{NSC_correction1} in Appendix \ref{app:3}). These two dEs could have formed with high angular momentum, yet low gas reservoir. Hence, their star formation was extinguished either intrinsically or by environmental processes in the group, a long time ago. Alternatively, before being accreted onto a group, these dE could have been quenched in a filament whose density was not high enough to affect their kinematics but adequate to quench their star formation \citep[see][]{2021Winkel}. Both explanations are in agreement with the high [$\alpha$/Fe] and low [M/H] we estimate for these two dEs (in case of VCC2019, after excluding its nuclear star cluster). 

We note that, using the STARLIGHT light-weighted ages for our sample dEs (Table \ref{Starlight_age_metal}) improves their consistency with the $\lambda_{\rm Re}$ - age relation, without
changing our interpretation.


\section{Conclusion}\label{Conclusion}

In this paper, we have investigated the stellar population properties of a sample of nine Virgo dEs that, according to \cite{Lisker2018}, have been accreted onto Virgo as gravitationally bound members of a massive galaxy group (M$_{\star}$/ M$_{\odot}$ $\sim$ 10$^{13}$) about 2-3 Gyr ago. To this effect, we have developed a four-step fitting routine of spectral indices based on a $\chi^{2}$ minimization approach, where  the observed indices of each galaxy are compared to predictions of SSP models with different age, [M/H], and [$\alpha$/Fe]. We have applied our fitting routine to
the integrated spectra of our sample of dEs, obtained from  their MUSE data cubes. Furthermore, we performed full-spectrum fitting (using the STARLIGHT code) to investigate the star formation history of the dEs in our sample. We have compared our results with those for dEs in the Virgo and Coma clusters spanning the same range of galaxy luminosity. The main results of our analysis can be summarized as follows:

\begin{itemize}
    \item Our dEs fall within the age range of 1.85 to 10.50 Gyr, the metallicity range of -0.77 $<$ [M/H] [dex]$<$ -0.18 and $\alpha$ abundance range of 0.20 $<$ [$\alpha$/Fe] [dex] $<$ 0.43. We obtain a median age of 6.8 $\pm$ 1.1 Gyr, median [M/H] of -0.59 $\pm$ 0.05 dex, and median [$\alpha$/Fe] of 0.26 $\pm$ 0.02 dex.
    \item We find our sample dEs to be significantly more $\alpha$-enhanced (at the 8$\sigma$ level) and metal-poorer (at the 2-3 $\sigma$ level) than equally-massive dEs of the Virgo and Coma clusters, at similar (or even larger) infall time. 
    \item Six out of nine dEs in our sample show significant peaks in their stellar light and mass fraction distributions between 1 to 4 Gyr ago, at or just after their accretion onto Virgo. These peaks correspond to a 5 to 30 percent mass growth, and point to a recent and short episode
    of star formation, possibly triggered by ram pressure in Virgo. Alternatively it might be due to the intrinsic star formation activity of these galaxies.
    We speculate that recent and rather short episodes of star formation might be (partly) responsible for the  high [$\alpha$/Fe] abundance ratios in these six dEs. This is consistent with the results of \cite{2011deLaRosa}, 
   who showed that galaxies with a short period of star formation (i.e., $\lesssim$ 2 Gyr) attain higher [$\alpha$/Fe] values. In the first few tens million years, type-II SNe enrich the interstellar medium (ISM) of $\alpha$-elements. If star formation lasts less than 1 Gyr (before Type-Ia SNe start to pollute the ISM with Fe), the newly 
  formed stellar population is more enriched in $\alpha$-elements than Fe, and exhibits high [$\alpha$/Fe]. If this population is the most recently formed and contributes 
  most of the galaxy light (as the star formation histories in Fig. \ref{ligh-mass-fractions} seem to indicate), then our measurements would actually reflect its [$\alpha$/Fe]. However, another possible explanation is that 
    the [$\alpha$/Fe] ratios might mostly arise from the old stellar component in these galaxies, whose star formation activity was quenched on relatively shorter timescales as the result of pre-processing in the host galaxy group, prior infall into Virgo.
    \item Our results indicate that the remaining three dEs in our sample (namely VCC0794, VCC1896 and VCC2019) were accreted onto Virgo already quenched. We suggest that their previous host group extinguished their star formation activity before their infall  into Virgo. Their relatively high [$\alpha$/Fe] ratios might thus
    be the result of quenching due to pre-processing in their previous parent halo.

\end{itemize}

In this work, we show that knowing the average infall times and star formation histories of recently accreted dEs onto Virgo may allow us to identify galaxies quenched by their previous host halo, thus providing direct proof of group pre-processing. Our results suggest that the combined effect of pre-processing in galaxy groups and environmental effects acting in the early phases of accretion onto a cluster are key drivers of the stellar population properties of low-mass galaxies. In a future work, we will further explore this scenario by analyzing the radially-resolved stellar population properties of our sample of dEs.

\section*{Acknowledgements}
We acknowledge financial support from the European Union's Horizon 2020 research and innovation program under the Marie Sklodowska-Curie grant agreement no. 721463 to the SUNDIAL ITN network. B.B. acknowledges the support of the International Max Planck Research School (IMPRS) for Astronomy and Cosmic Physics at the University of Heidelberg.
J.~F-B   and F.~LB acknowledges support through the RAVET project by the grant PID2019-107427GB-C32 from the Spanish Ministry of Science, Innovation and Universities (MCIU), and through the IAC project TRACES, which is partially supported through the state budget and the regional budget of the Consejer\'ia de Econom\'ia, Industria, Comercio y Conocimiento of the Canary Islands Autonomous Community. GvdV acknowledges funding from the European Research Council (ERC) under the European Union's Horizon 2020 research and innovation programme under grant agreement No 724857 (Consolidator Grant ArcheoDyn). The MUSE data cubes used in this work are available in the ESO archive under the ESO programme IDs 098.B-0619 and 0100.B-0573. This research is partially based on data from the MILES project.

\section*{Data Availability}
Data and results directly referring to content and figures
of this publication are available upon request from the corresponding
author. 
 


\bibliographystyle{mnras}
\bibliography{Main} 



\appendix

\section{List of Index band passes }\label{app:1}

In Table \ref{Index banpass} we provide the list of Lick indices used in this study. 
\begin{table*}
\caption{\label{Index banpass} List of indices used in this study}
\centering
\begin{tabular}{c c c c c}
\hline
Index & Blue bandpass [\AA] & Central bandpass [\AA] & Red bandpass [\AA] & Reference \\
\hline
\hline
H$\beta$ & 4827.875-4847.875 & 4847.875-4876.625&4876.625- 4891.625& \citep{1998Trager}\\
H$\beta_{o}$ &4821.175-4838.404& 4839.275-4877.097 & 4897.445- 4915.845& \citep{2009Cervantes}\\
Fe5015 & 4946.500-4977.750 & 4977.750-5054.000& 5054.000- 5065.2500& \citep{1998Trager}\\
Mgb5177 & 5142.625 - 5161.375 & 5160.125-5192.625& 5191.375- 5206.375& \citep{1998Trager} \\
Fe5270 &5233.150-5248.150 & 5245.650-5285.650& 5285.650- 5318.150 &\citep{1998Trager}\\
Fe5335& 5304.625- 5315.875 & 5312.125-5352.125& 5353.375- 5363.375& \citep{1998Trager}\\
\hline
\end{tabular}\\
The columns show: index’s name, its corresponding blue pseudo-continuum, main absorption feature, red pseudo-continuum, and the reference where the index was defined. \\
\end{table*}

\section{Lick indices vs STARLIGHT fitting}\label{app:2}

\begin{figure}
    \centering
	\includegraphics[scale=0.5]{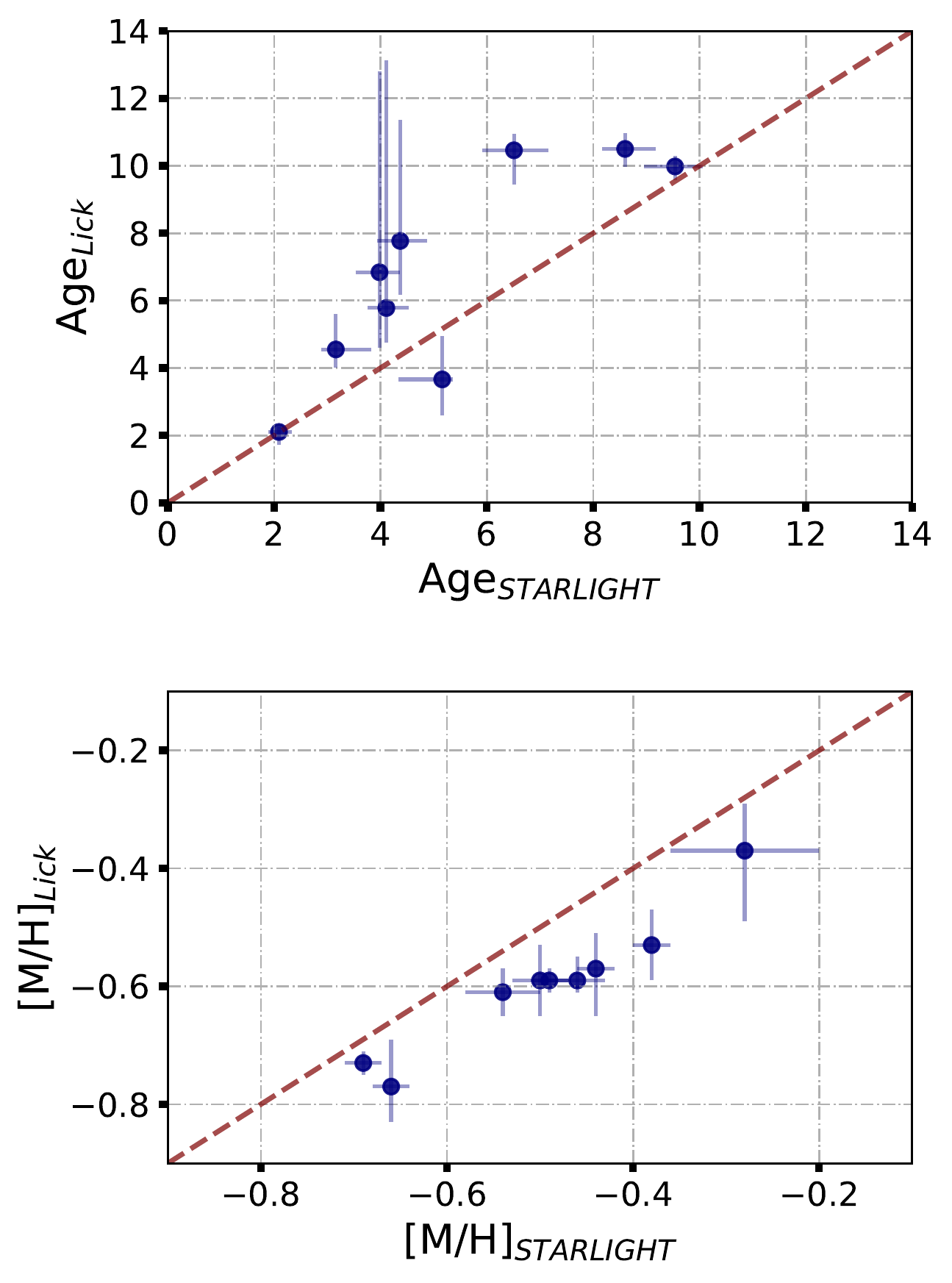}
	\caption{\textit{Top panel}: Comparing light-weighted age values obtained from STARLIGHT full spectrum fitting (x-axis) with those derived through Lick indices fitting (y-axis). The data points represent dEs in our sample. \textit{Bottom panel}: Same comparison as the top panel but for light-weighted [M/H] values.} 
	\label{stellar_fits_compare}
\end{figure}

In Fig.\ref{stellar_fits_compare} we compare the results of STARLIGHT (Table \ref{Starlight_age_metal}) with those from the Lick indices fitting (Table \ref{Integrated values}). In the top panel of this figure, we compare the light-weighted mean ages obtained from STARLIGHT (on the x-axis) and the average ages derived by fitting over Lick indices (on the y-axis). A similar comparison, but for light-weighted mean [M/H], is presented in the bottom panel. A KS-test applied to these two sets shows that
the two fitting methods are generally consistent, with
p-value = 0.35 and 0.12 for age and metallicity, respectively).

\ref{stellar_fits_compare} shows that STARLIGHT generally underestimates age and overestimates metallicity with respect to Lick indices. This can partly be due to systematic and fundamental differences between the two fitting methods. STARLIGHT considers a mixture of SSP models to construct the best fitting model. Thus, for galaxies with multiple epochs of star formation, STARLIGHT provides a more realistic approach than the Lick indices fitting based on one SSP. Precisely, the light-weighted age and [M/H] in Table \ref{Starlight_age_metal} are derived with STARLIGHT by accounting for the possible contribution of both young and old SSPs in the observed spectrum, whereas in Table \ref{Integrated values}, we base our Lick indices analysis solely on one SSP without simultaneously considering different SSP components. 

In Section \ref{method:STARLIGHT}, we state that STARLIGHT accepts only a maximum 
number of 300 SSPs, while for the Lick indices fitting, such limitation does not exist (we use $\sim$ 2,410,000 SSP models). Hence, STARLIGHT determines the best-fitting spectrum based on a limited model space that has uneven spacing in the [M/H] axis and a spacing of 1 Gyr in the age axis. This may affect the contribution of each SSP model to the best fitting spectrum, thus the final average values of age and [M/H]. It should also be noted that the uncertainties due to model degeneracies (particularly for models older than 6 Gyr) is  still relevant for the results of the full-spectrum fitting. In addition, the results of the Lick indices fitting noticeably suffer from the well-known issue of the “age-metallicity degeneracy”. According to \cite{2011MNRAS.415..709S}, this issue is less relevant for full-spectrum fitting, yet it should be taken into account while interpreting the present discrepancy between the results of these two approaches. 

\section{The effects of the nuclear star cluster in VCC0990 and VCC2019}\label{app:3}

\begin{table*}
\caption{\label{The NSC test} Integrated light-/mass-weighted stellar population properties of dEs with NSC }
\centering
\begin{tabular}{c c c c c c}
\hline
Object & R$_{\rm NSC}$ & $<$Age$>$$_{\rm L}$ & $<$[M/H]$>$$_{\rm L}$ & $<$Age$>$$_{\rm M}$ & $<$[M/H]$>$$_{\rm M}$ \\
    & [arcsec] & [Gyr] & [dex] & [Gyr] & [dex] \\
\hline
\hline
VCC0990 & 3.2 & 5.76 $\pm$ 0.43 & -0.49 $\pm$ 0.11 & 6.54 $\pm$ 0.50 & -0.45 $\pm$ 0.10\\
VCC2019 & 3.9 & 8.27 $\pm$ 0.85 & -0.55 $\pm$ 0.15 & 8.48 $\pm$ 0.85 & -0.51 $\pm$ 0.13\\
\hline
\end{tabular}\\
The columns show: Columns are: name of the target, NSC radius, galaxy's light-weighted age, light-weighted metallicity, mass-weighted age, and mass-weighted metallicity. \\
\end{table*}

We investigate how the inferred star formation history, age, and [M/H], of VCC0990 and VCC2019 change after excluding their central nuclear star cluster (NSC). To reliably estimate the NSC size,
we first discard spaxels with S/N$<$3 and then sum each dE’s MUSE data cube along the wavelength axis, between 4750 to 5540 \AA. This is the exact wavelength range over which we performed STARLIGHT fitting in Section \ref{method:STARLIGHT}. We then stack all the rows of the resulting image to construct the light distribution of the entire galaxy, which we simultaneously fit with two Gaussian functions, one with a broad kernel for the diffuse galaxy, and the other with a narrow kernel for the galaxy's NSC. We adopt 
the FWHM of the best-fitting narrow Gaussian as the size of the NSC, whose value is reported in Table \ref{The NSC test} for each of the two galaxies.
We then exclude the spaxels in the central area of the galaxy as defined by its NSC radius, and
construct a new integrated MUSE spectrum for each galaxy.
We fit both spectra using STARLIGHT as described in 
Section \ref{method:STARLIGHT}, and report the resulting light- and mass-weighted age and [M/H] in Table \ref{The NSC test}.
The exclusion of most of the NSC emission from their integrated spectra turns VCC0990 and VCC2019 older (at the 1.7$\sigma$ level) and slightly
metal-poorer (at the 1$\sigma$ level).
As stated in Section \ref{res:3} the presence of a young NSC in the core of VCC0990 and VCC2019 leverages the contribution of young SSPs in the STARLIGHT best fit. 

\begin{figure*}
    \centering
	\includegraphics[scale=0.5]{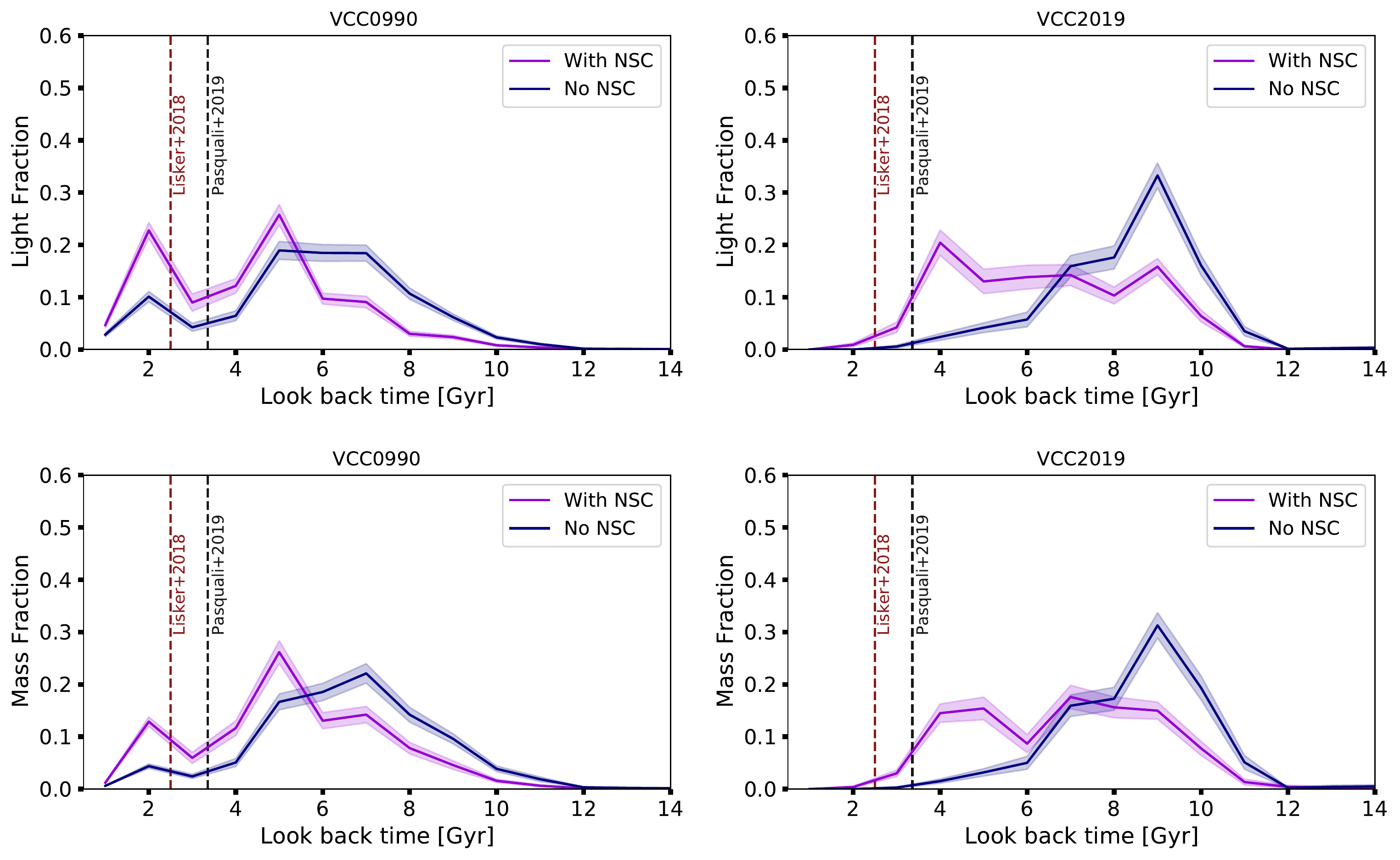}
	\caption{Similar to Fig.\ref{ligh-mass-fractions} but only for VCC0990 (left panels) and VCC2019 (right panels). In each panel, the light and mass fractions before and after excluding the NSC are traced with purple and blue, respectively.}
	\label{NSC_correction1}
\end{figure*}

\begin{figure*}
    \centering
	\includegraphics[scale=0.5]{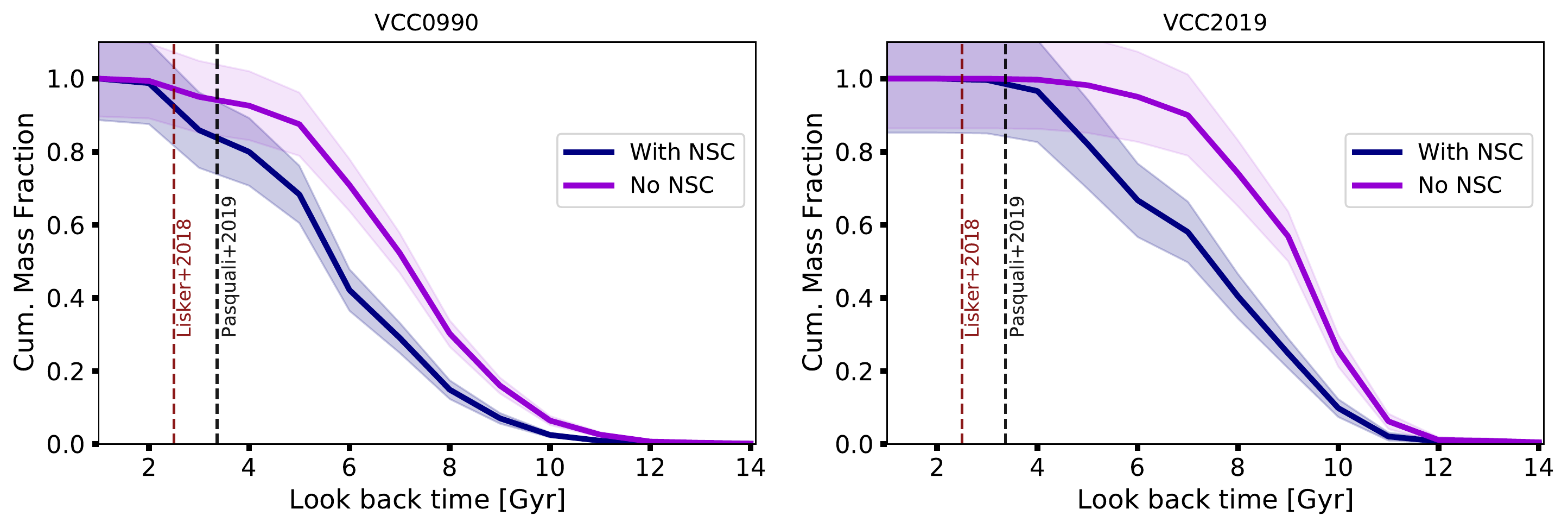}
	\caption{Similar to Fig.\ref{cumulative_prof} but only for VCC0990 (left panel) and VCC2019 (right panel). Similar to Fig.\ref{NSC_correction1} the light and mass fractions before and after excluding the NSC are traced with purple and blue, respectively.}
	\label{NSC_correction2}
\end{figure*}

The light and mass fractions of VCC0990 and VCC2019, before and after excluding their NSC, are traced in Fig.\ref{NSC_correction1} as a function of time. Their corresponding cumulative mass distributions are shown in Fig.\ref{NSC_correction2}. The peak at younger ages (i.e., at $\sim$ 2 Gyr for VCC0990 and $\sim$ 4 Gyr for VCC2019) in the star formation history of these two dEs is
noticeably reduced after excluding their NSC from their integrated spectra. This indicates that VCC2019 was quenched mostly before its accretion onto the Virgo cluster. According to  Fig.\ref{NSC_correction1} the star formation activity
of VCC0990 continued even after its accretion onto the Virgo cluster. Nevertheless, this phase of star formation corresponds to less than 10 percent stellar mass growth (as shown in Fig.\ref{NSC_correction2}), and it might be due to possible 
residual NSC emission in the galaxy's integrated spectrum.

\bsp	
\label{lastpage}
\end{document}